\documentclass[aps,pre,twocolumn,superscriptaddress,showpacs]{revtex4-1}
\usepackage{graphicx,amsbsy,amsmath,epsfig,natbib,float}
\usepackage{amssymb,latexsym,wasysym,pifont,mathrsfs}
\usepackage{comment,verbatim,multirow,array,sidecap,color}
\usepackage{lineno,setspace,soul,cancel,enumerate,amsmath,mathtools}
\usepackage[normalem]{ulem}
\begin{document}

\title{The role of annealing in determining the yielding behavior of glasses under cyclic shear deformation}

\author{Himangsu Bhaumik}
\affiliation{Jawaharlal Nehru Center for Advanced Scientific Research, Jakkur Campus, Bengaluru 560064, India.}
\author{Giuseppe Foffi}
\affiliation{Laboratoire de Physique des Solides, CNRS, Universit\'e Paris-Sud, Universit\'e Paris-Saclay, 91405 Orsay, France}
\author{Srikanth Sastry}
\email{sastry@jncasr.ac.in}
\affiliation{Jawaharlal Nehru Center for Advanced Scientific Research, Jakkur Campus, Bengaluru 560064, India.}






\begin{abstract}
Yielding behavior in amorphous solids has been investigated in
computer simulations employing uniform and cyclic shear
deformation. Recent results characterise yielding as a discontinuous
transition, with the degree of annealing of glasses being a
significant parameter. Under uniform shear, discontinuous changes in
stresses at yielding occur in the high annealing regime, separated
from the poor annealing regime in which yielding is gradual. In cyclic
shear simulations, relatively poorly annealed glasses become
progressively better annealed as the yielding point is approached,
with a relatively modest but clear discontinuous change at
yielding. To understand better the role of annealing on yielding
characteristics, we perform athermal quasistaic cyclic shear
simulations of glasses prepared with a wide range of annealing in two
qualitatively different systems -- a model of silica (a network
glass), and an atomic binary mixture glass.  Two strikingly different
regimes of behavior emerge: Energies of poorly annealed samples evolve
towards a unique threshold energy as the strain amplitude increases,
before yielding takes place. Well annealed samples, in contrast, show
no significant energy change with strain amplitude till they yield,
accompanied by discontinuous energy changes that increase with the
degree of annealing.  Significantly, the threshold energy for both
systems correspond to dynamical crossover temperatures associated with
changes in the character of the energy landscape sampled by glass
forming liquids. Uniform shear simulations support the recently
discussed scenario of a random critical point separating ductile and
brittle yielding, which our results now associate with dynamical
crossover temperatures in the corresponding liquids.

\end{abstract}


\maketitle 

The response of structural materials to applied stresses is
of fundamental importance in determining their utility. Many
structural materials are amorphous solids, with molecular glasses
being a predominant example. In addition to specific types of
molecular glasses, many other amorphous solids are the subject of
ongoing fundamental and applied research investigations, such as
colloidal suspensions, foams, emulsions and granular packings
\cite{ramamurty,falkARCMP11,Bonn2017c,Nicolas2018}. Upon increasing
deformation or applied stresses, the mechanical response of such
amorphous solids generically changes from elastic solid-like response
at small deformations, to elasto-plastic flow at large
deformations. But the transition between these two regimes occurs in
diverse ways. For typical molecular glasses (window glass or silica,
and metallic glasses), yielding is a sudden, sharp event,
characterised as brittle failure. For many {\it soft} solids, the
transition between the elastic and plastic flow regimes is gradual
(ductile), with no sharply defined transition point between
them. Prominent among the challenges in understanding yielding is the
rationalisation of this diversity. A number of theoretical and
computational investigations have recently addressed the nature of the
yielding transition
\cite{wisitsorasak2012strength,wolynesshearbanding,Lin2014,Fiocco2013,regev2015reversibility,RainoneetalPRL2015,Urbani2017b,Jin2018,jaiswalPRL2016,parisiPNAS2017,procacciaetal2017,kawasakiPRE16,leishangthem2017,Priezjev2013,PRIEZJEV2018,ozawaPNAS2018,Popovic2018a,parmarPRX2019},
many of which support the notion that yielding must be understood as a
spinodal limit, at which system-spanning plastic deformation occurs
discontinuously. The nature of such a limit (whether it is {\it
  critical} or not, and whether correlated plastic rearrangements or
{\it avalanches} diverge upon approaching it, {\it etc.}) are matters
of ongoing debate. Computational investigations have typically focused
on the limit of athermal quastistatic (AQS) deformation of atomistic
models, employing both uniform
\cite{shi2005strain,shi2007,Jin2018,jaiswalPRL2016,parisiPNAS2017,procacciaetal2017,ozawaPNAS2018}
or cyclic
\cite{Fiocco2013,regev2015reversibility,kawasakiPRE16,leishangthem2017,PRIEZJEV2018,parmarPRX2019}
shear deformation, and simulations of elasto-plastic models
\cite{Lin2014,ozawaPNAS2018,Popovic2018a}. Although the role of the
degree of annealing in determining the mechanical response of glasses
has been studied from early on \cite{shi2005strain,shi2007}, the focus
has been on features such as strain localisation, and not the nature
of the yielding transition itself.  Recently, Ozawa {\it et
  al.},\cite{ozawaPNAS2018} investigated the dependence on annealing
of the nature of yielding by considering uniform shear simulations of
glasses prepared over a wide range of temperatures. Based on their
numerical results and accompanying theoretical analysis, they
identified two distinct regimes of annealing, one of high annealing in
which yielding is discontinuous, and the other of poor where it is
gradual, and argued that a random critical point (of the class
described by the random field Ising model (RFIM)) separates the two
regimes at a critical value of annealing (or disorder). Investigations
employing cyclic shear deformation, with both poorly annealed and
moderately well annealed glasses \cite{leishangthem2017,parmarPRX2019}
reveal an apparently different picture. For amplitudes of strain below
a critical value, repeated cyclic shear results in annealing the
glasses, with the degree of annealing increasing with an increase in
the strain amplitude. Yielding (more precisely, {\it fatigue} failure)
occurs discontinuously at a critical strain value, with a jump in the
peak stress, the emergence of diffusive motion of particles, and shear
banding
\cite{regev2015reversibility,kawasakiPRE16,leishangthem2017,PRIEZJEV2018,parmarPRX2019}.

An important question raised by the above results is whether uniform
and cyclic shear protocols lead to qualitatively different
manifestations of yielding behavior. Additionally, one may ask how
general the results in earlier work
\cite{ozawaPNAS2018,leishangthem2017,parmarPRX2019} are, since the
studied systems have all been models of atomic glasses. Finally, if a
threshold degree of annealing or disorder does robustly separate
brittle and ductile yielding behaviors, one may ask whether we can
attribute any physical significance to the critical annealing. We
address these questions in the present work.  We consider cyclic shear
simulations of a model atomic glass studied earlier (the Kob-Andersen
binary mixture Lennard-Jones model (KA BMLJ))
\cite{leishangthem2017,parmarPRX2019}, but over a much broader degree
of annealing. As an important example of a distinct class of glasses
with which to study the generality of the aforementioned results, we
study silica, a network forming glass. From the analysis of these two
models, we conclude that cyclic shear indeed reveals the transition
from ductile to brittle behavior, consistently with uniform shear
investigations, revealing new and interesting characteristics of the
transition. Finally, our results reveal an interesting physical
significance to the critical degree of annealing -- we find that the
critical degree of disorder corresponds to well studied dynamical
crossovers in the studied systems, the fragile-strong crossover in
silica, and the mode coupling temperature in the KA BMLJ model.

For both the model systems we study, the glasses we subject to AQS
deformation are generated by local energy minimization of simulated
instantaneous liquid configurations. Based on the well studied
relationship between temperature and energies of typical minima or
{\it inherent structures} (IS) ({\it e. g.} , \cite{SastryNature1998};
{\it Appendix}, Fig. A1), the degree of annealing of the glasses is
equivalently specified by the potential energy of the inherent
structures, or the temperature of the equilibrium liquid from which
they are obtained.

We simulate the BKS model \cite{beestPRL90} of silica as implemented
in Voivod {\em et al.}  \cite{voivodPRE04a} for a temperature range
from $T=2500$K to $6000$K, at density $\rho=2.8$gm/cm$^3$ using
$N=1728$ ions. Arising from its tetrahedral local geometry, silica
exhibits a liquid-liquid phase transition
\cite{voivodPRE00,chenJCP17}, and an associated fragile to strong
crossover \cite{horbachPRB99,voivodNAT01,voivodPRE04a,heuerPRL04},
which occurs around $T = 3100 K$ for $\rho=2.8$gm/cm$^3$, within the
range of temperatures we simulate.

The KA BMLJ is simulated at reduced density $\rho = 1.2$, for $N =
4000$, for a temperature range from $T = 0.435$ to $T = 1.5$,
corresponding to IS energies from $E_{IS}=-6.89$ to around $ -7.0$. We
also study inherent structures obtained through a finite temperature,
finite shear rate cyclic shear annealing procedure reported in
\cite{das2018annealing}, which extends the range of energies down to
$E_{IS} = -7.07$ (to be compared to the extrapolated Kauzmann energy
of $E^{K}_{IS} = -7.15$ \cite{das2018annealing}). We employ the same
procedure for system sizes $N = 2000, 8000, 16000, 32000, 64000$, to
obtain glasses with lowest energy $E_{IS} = -7.05$, for performing a
system size analysis.

These initial inherent structures are subjected to athermal
quasi-static shearing (AQS) protocol involving: (i) Affine
transformation by small strain increments of $d\gamma = 2\times
10^{-4}$ in the $xz$-plane ($x^\prime\to x+d\gamma z$, $y^\prime\to
y$, $z^\prime\to z$). (ii) Energy minimization. The strain $\gamma$ is
varied cyclically: $0 \to \gamma_{\rm {max}}\to -\gamma_{\rm{max}}\to
0$ (or uniformly, for part of the study). Repeated deformation cycles
for a given $\gamma_{\rm{max}}$ result in the glasses reaching steady
states wherein the properties on average do not change.  Further
details regarding simulations may be found in {\it Appendix} Sec. {1} and {\it 2}).

\subsection*{Cyclic Shear Deformation of BKS Silica} 

\begin{figure*}[t]
  \centerline{
       \includegraphics[width=.45\linewidth]{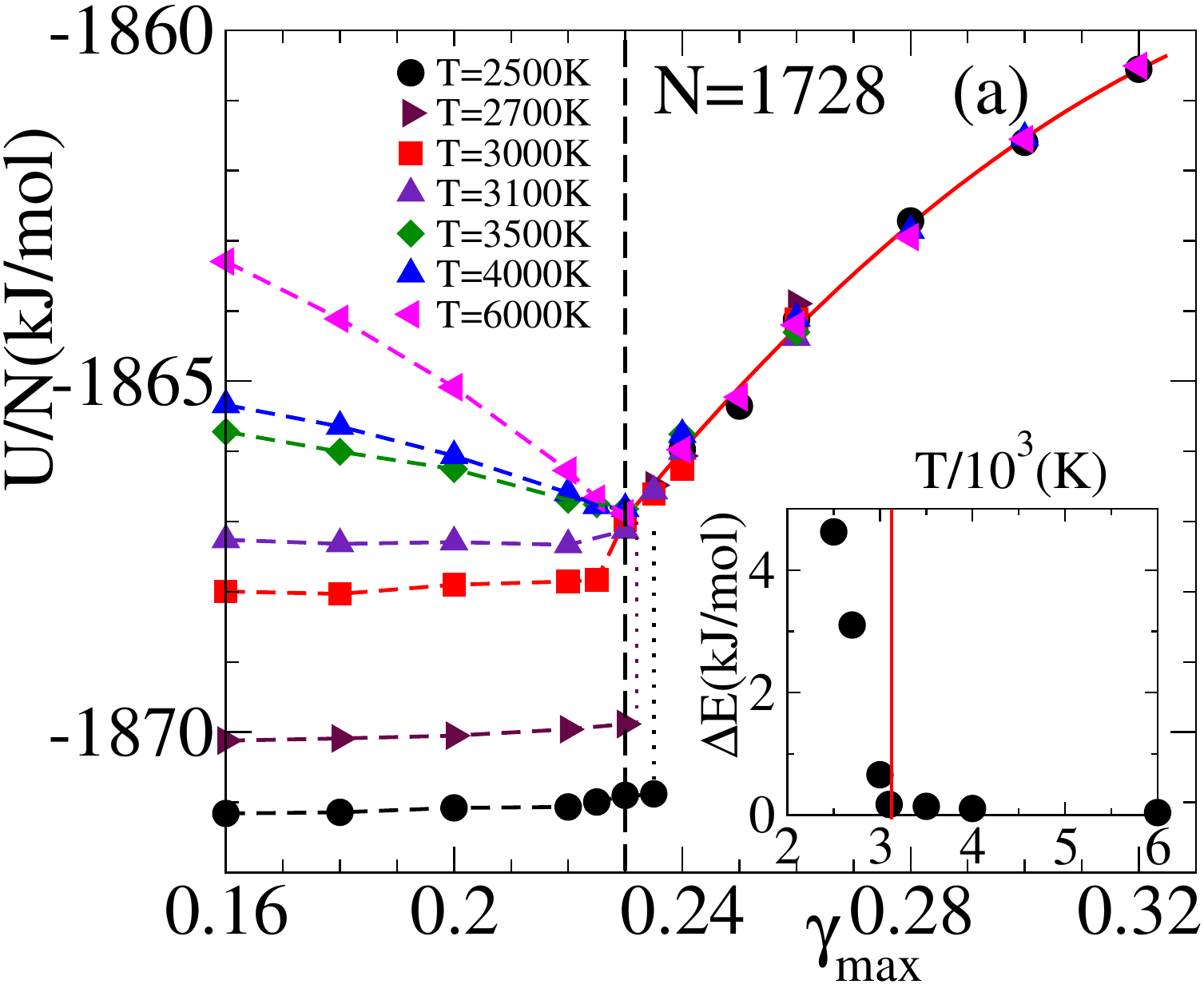}
  \includegraphics[width=.42\linewidth]{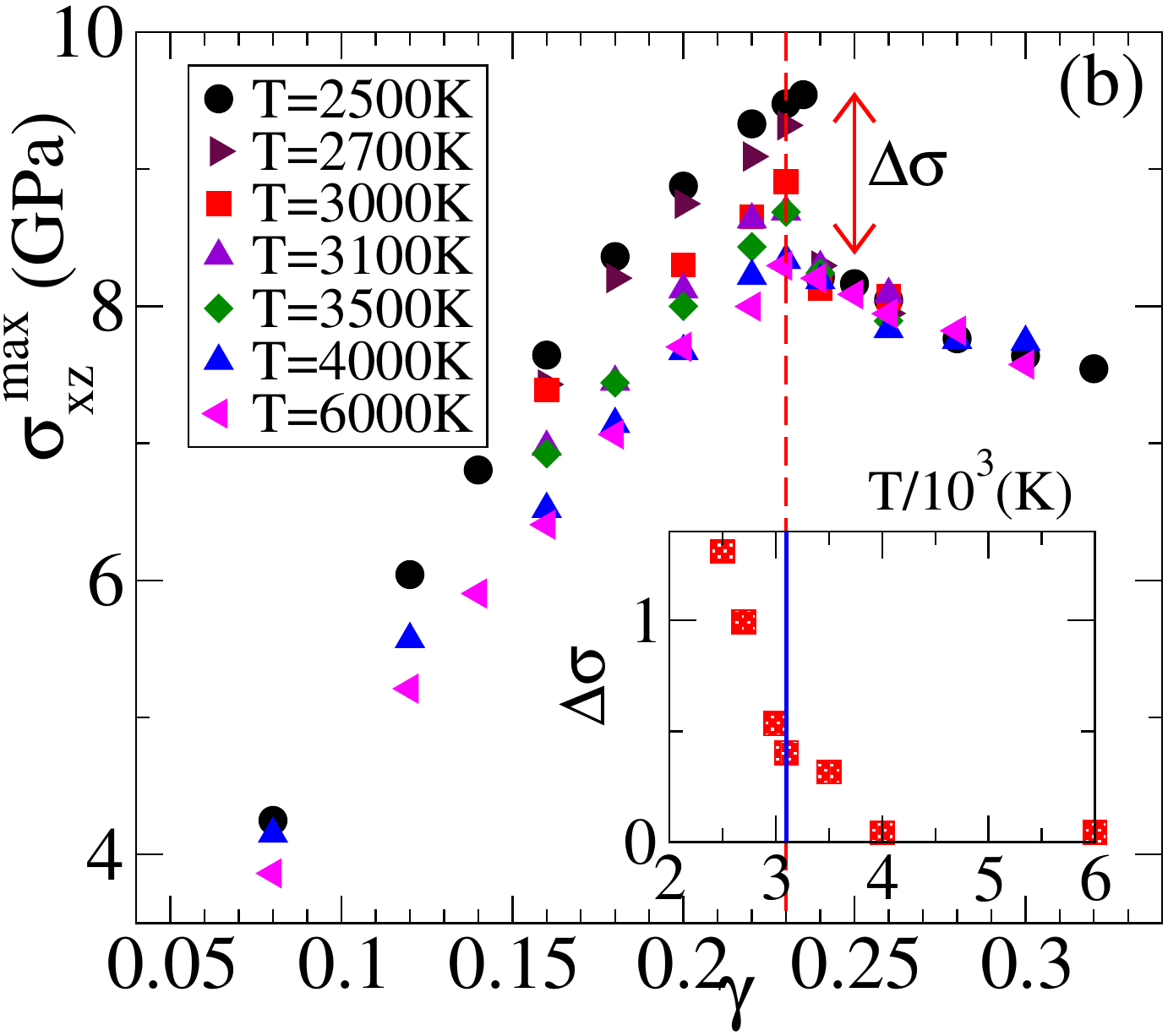}
  }
  \centerline{
       \includegraphics[width=.44\linewidth]{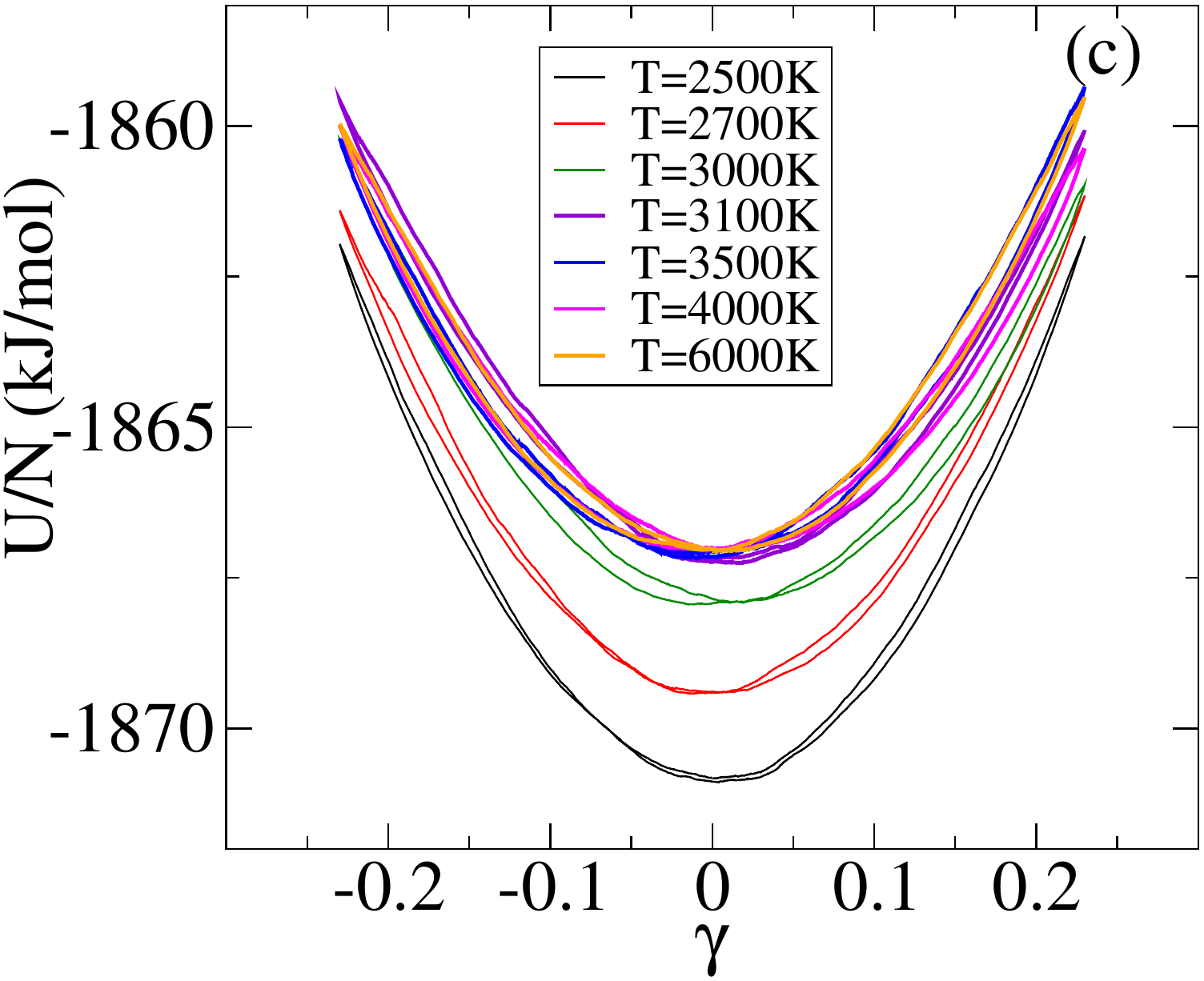}
  \includegraphics[width=.44\linewidth]{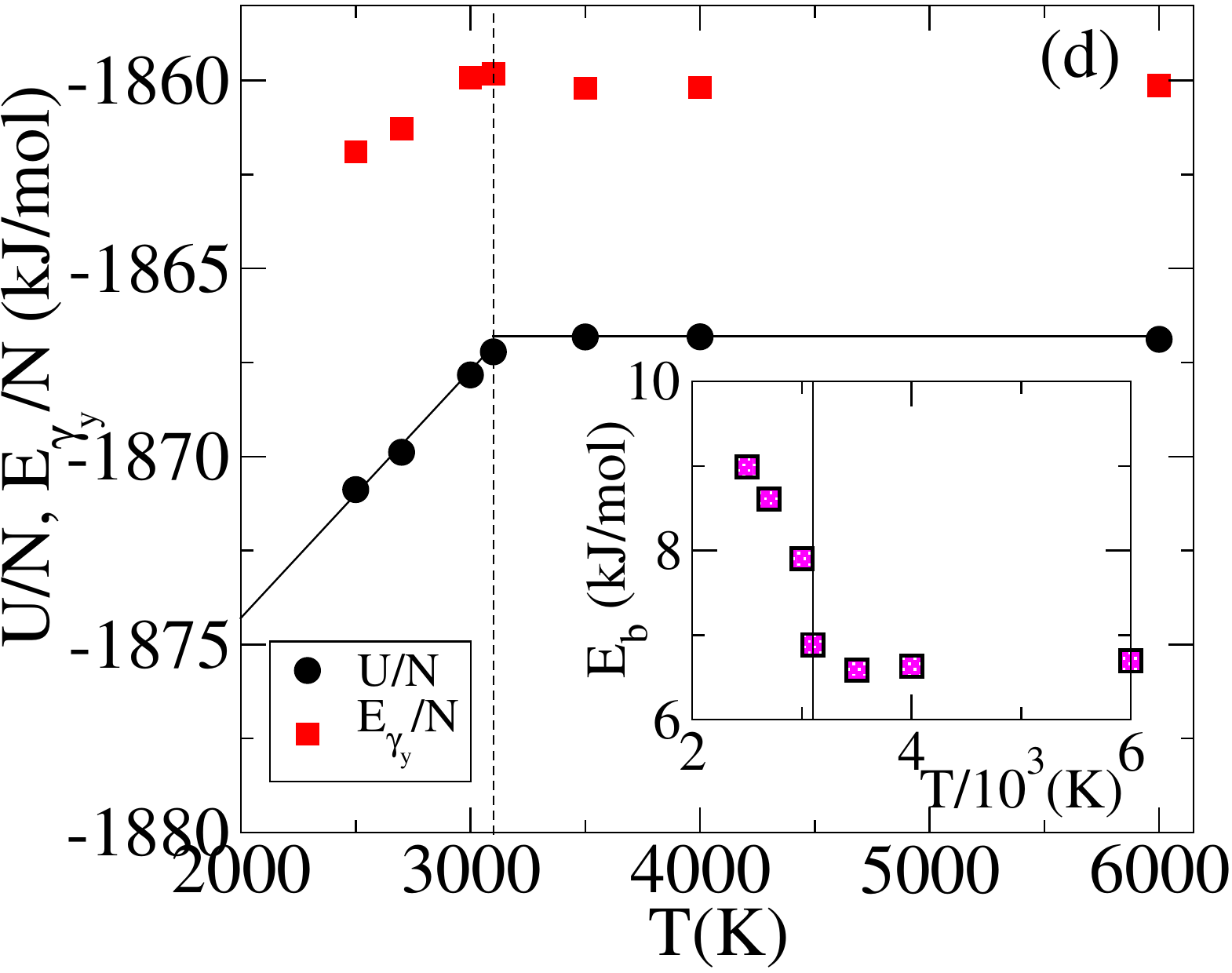}
  }
    \caption{\label{fig_silica} {\bf BKS Silica under cyclic shear:}
      ($N=1728$) (a) The steady state energies (stroboscopic) are
      plotted against $\gamma_{max}$ for different temperatures. Inset
      shows the jump in energy $\Delta E$ at the yielding point. (b)
      The maximum stress value is plotted against strain amplitude for
      different temperatures. The magnitude of jump in the maximum
      stress $\Delta \sigma$ at the critical strain is plotted against
      $T$ in the inset. (c) Plot of the potential energy versus
      $\gamma$ in the steady state for different temperatures, at the
      yield strain amplitude. (d) Plot of energies $E_{\gamma_{y}}/N$
      at $\gamma = \gamma_{max}=\gamma_y$, and $U/N$ (at $\gamma=0$)
      for the yield amplitude $\gamma_{max} = \gamma_{y}$ as a
      function of temperature. Inset shows the difference between
      $E_{\gamma_{y}}/N$ and $U/N$ {\it vs.} $T$. Vertical lines in
      the insets are at $T = 3100 K$.}
\end{figure*}
We first describe results from cyclic shear deformation of silica,
which are shown in Fig. \ref{fig_silica}. In Fig.
\ref{fig_silica}(a), we show for a wide range of temperatures the
steady state (zero strain) energy $U$ obtained in the limit of large
numbers of cycles. The data shown are obtained by stretched
exponential fits of the stroboscopic (at the end of each cycle, strain
$\gamma = 0$) values of the energy as a function of the accumulated
strain $\gamma_{acc} = 4 \times \gamma_{max} \times N_{cyc}$ where
$N_{cyc}$ is the number of elapsed cycles (Additional analysis data
pertaining to Fig.  \ref{fig_silica} is found in Fig.s A2-A4 of {\it
  Appendix}). The data for the highest three temperatures, $T = 6000 K, 4000
K, 3500 K$, reveal the yielding strain $\gamma_{y} = 0.23$ as a point
of minimum energy, consistently with previous work
\cite{leishangthem2017,parmarPRX2019}, but with a feature that was not
clear in earlier work -- as the yield strain amplitude $ \gamma_{max}
= \gamma_{y}$ is approached, the energy of all the glasses
corresponding to this temperature range converge to a common energy
value of $-1867 kJ/mol$, corresponding to a temperature of $T_{th} =
3100 K$. At lower temperatures, a surprising new feature appears -- to
a very good extent, the energies of the glasses do not change with
$\gamma_{max}$, and at the yielding point (with a yield strain that
increases mildly as the glass energy decreases), the energy shows a
discontinuous jump, whose size increases as one goes to lower
temperatures, as shown in the inset of Fig.  \ref{fig_silica}(a). Both
the relative lack of annealing during cyclic shear, and the increasing
size of the energy jump upon yielding, are new features. Next we
consider the variation of maximum stress $\sigma_{xz}^{max}$ with
strain amplitude for the different cases, and find (as shown in Fig.
\ref{fig_silica}(b)) that the size of the stress jump at yielding
increases with decreasing temperature/degree of annealing. At the
highest two temperatures, our results indicate that the stress drop at
yielding are very small, but increase rapidly below $T = 3100K$.  In
Fig.  \ref{fig_silica}(c) we show energies through the cycle for the
studied temperatures in the steady state, at the largest strain
amplitude below yield, identified as the yield strain amplitude
$\gamma_{y}$. These data reveal that (unlike the KA BMLJ; see below)
observable plasticity remains even in well annealed samples in the
steady state, a feature that merits further analysis. We extract the
energy at zero strain, and at $\gamma = \gamma_{y}$ from this data,
shown in Fig.  \ref{fig_silica}(d), which clearly indicate that both
the zero strain energy, and the energy at the yield strain, decrease
below $T = 3100 K$, but they do so differently. The difference between
the two, which may be thought of as an energy barrier to yielding,
increases with decrease of temperature below $T = 3100 K$ (shown in
Fig.  \ref{fig_silica}(d) inset).

\subsection*{Cyclic shear deformation of the Kob-Andersen Binary Mixture Lennard-Jones Model}

\begin{figure*}[t]
\centerline{
     \includegraphics[width=.45\linewidth]{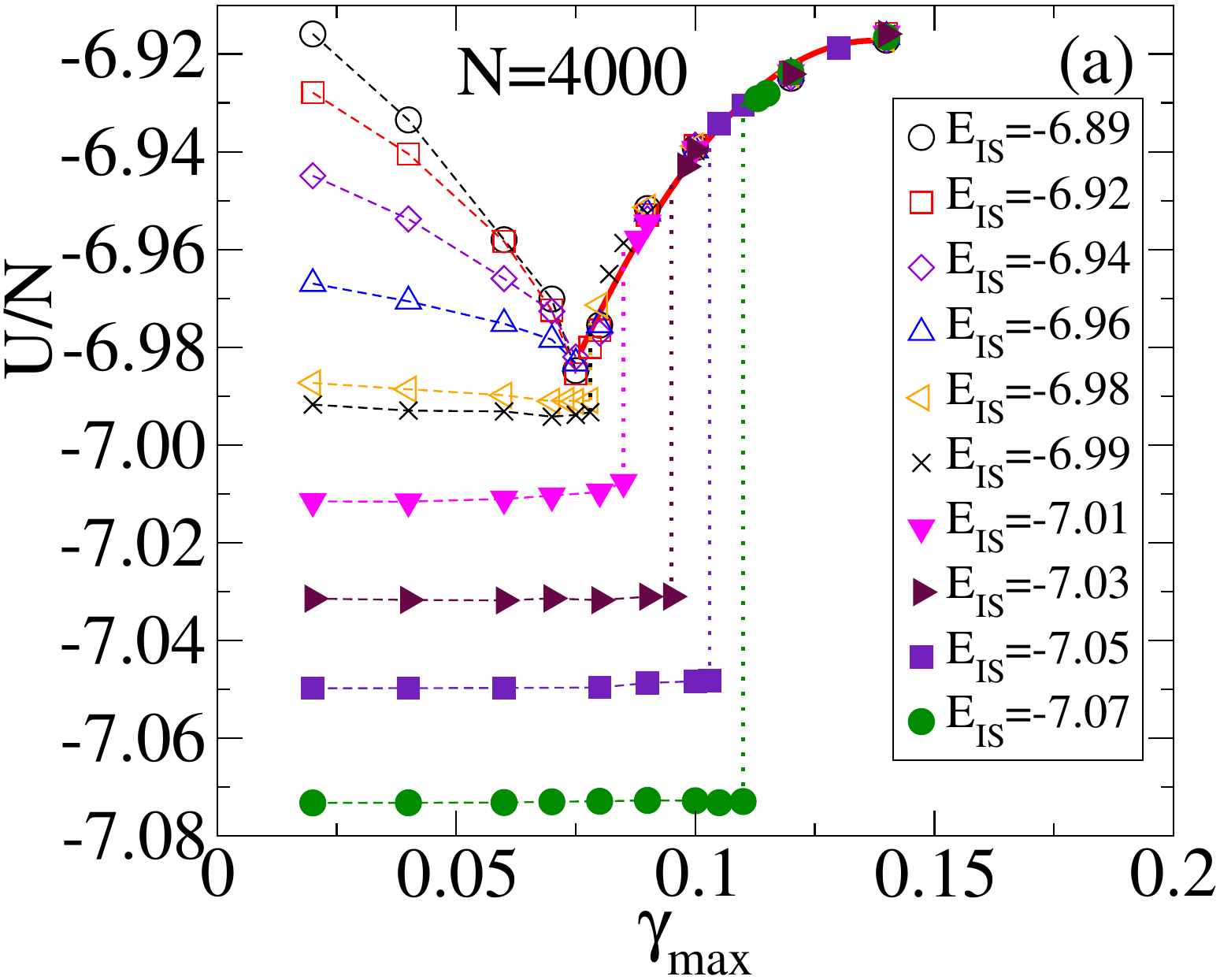}
   \includegraphics[width=.44\linewidth]{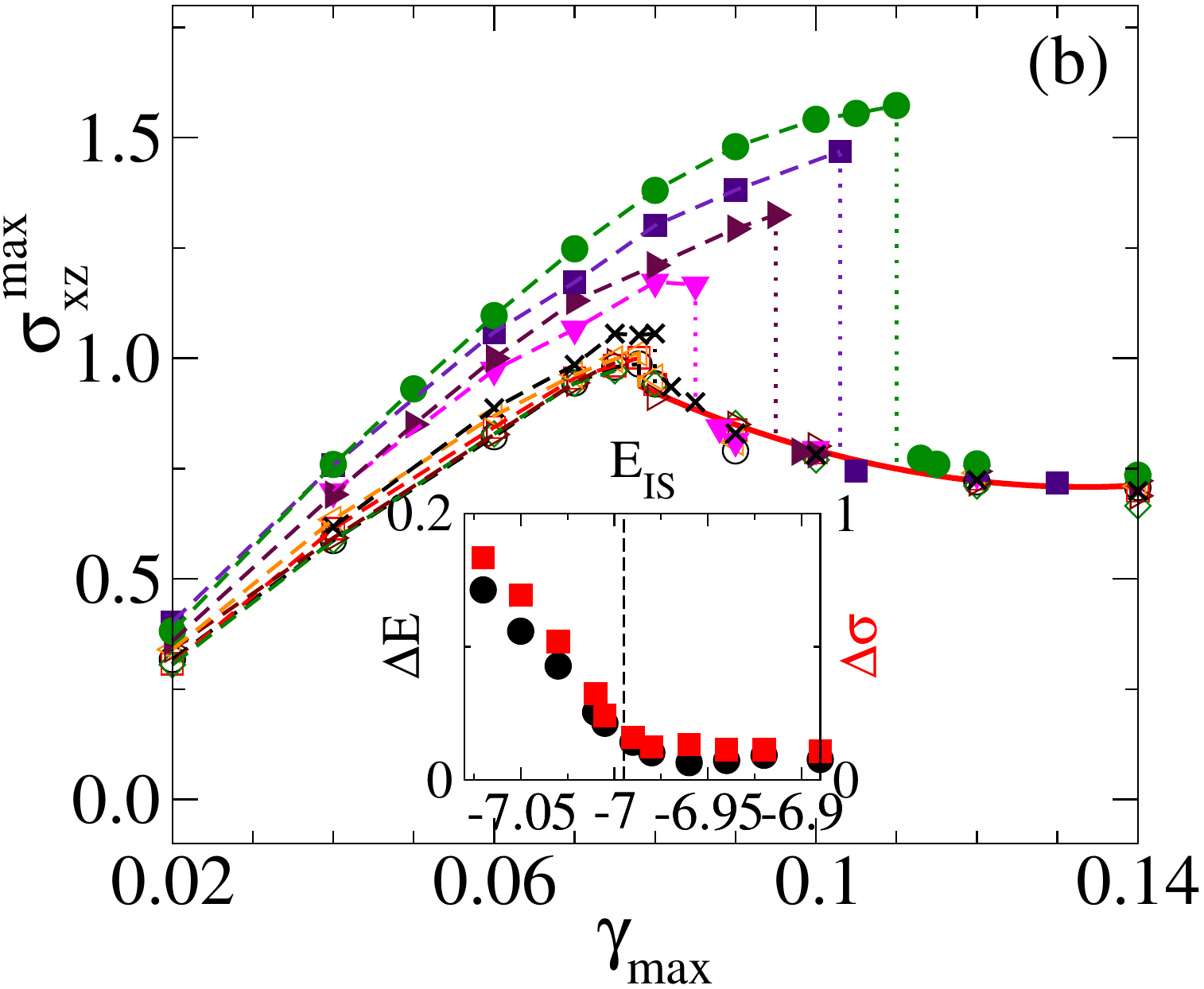}
   }
   \centerline{
  \includegraphics[width=.43\linewidth]{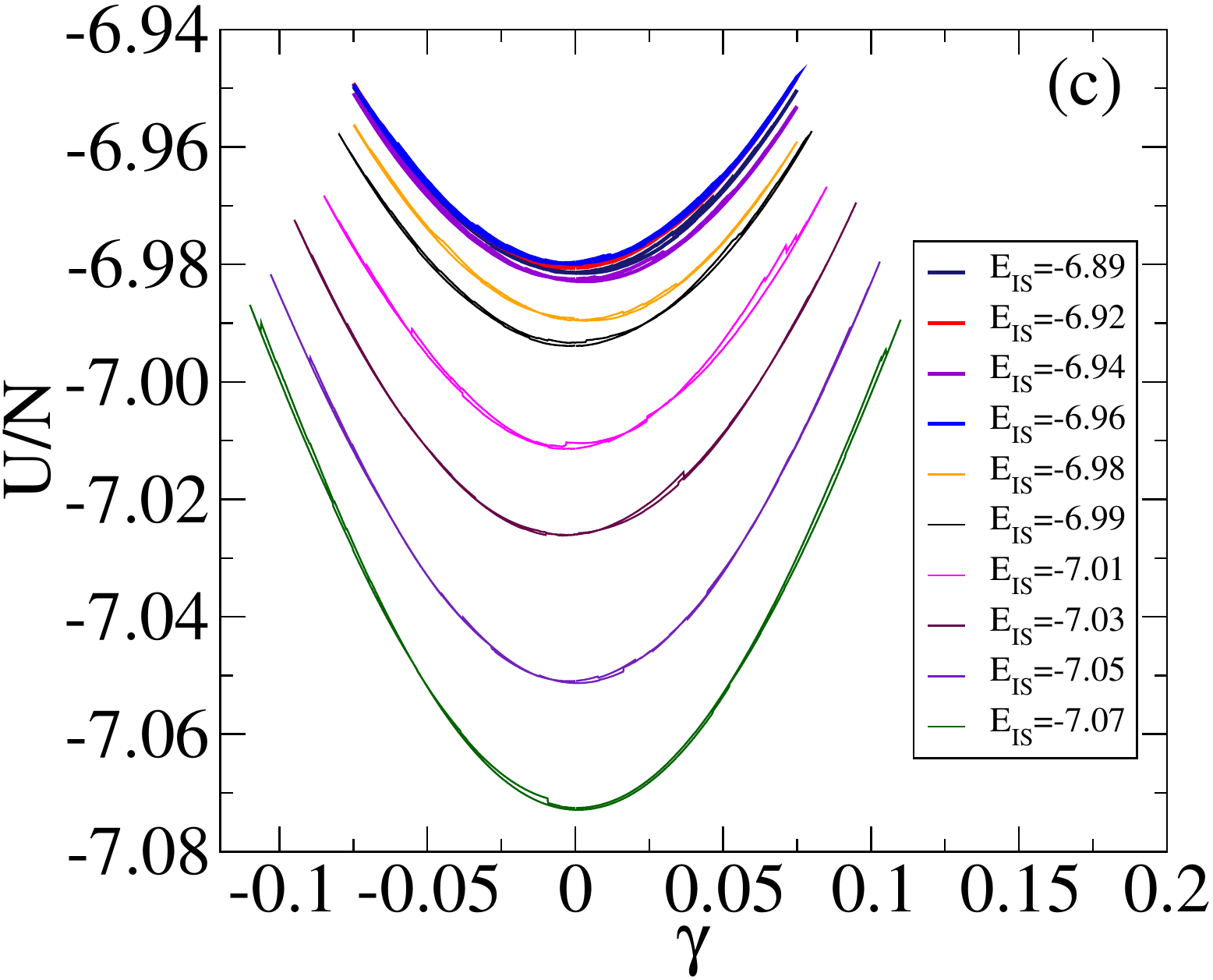}
  \includegraphics[width=.43\linewidth]{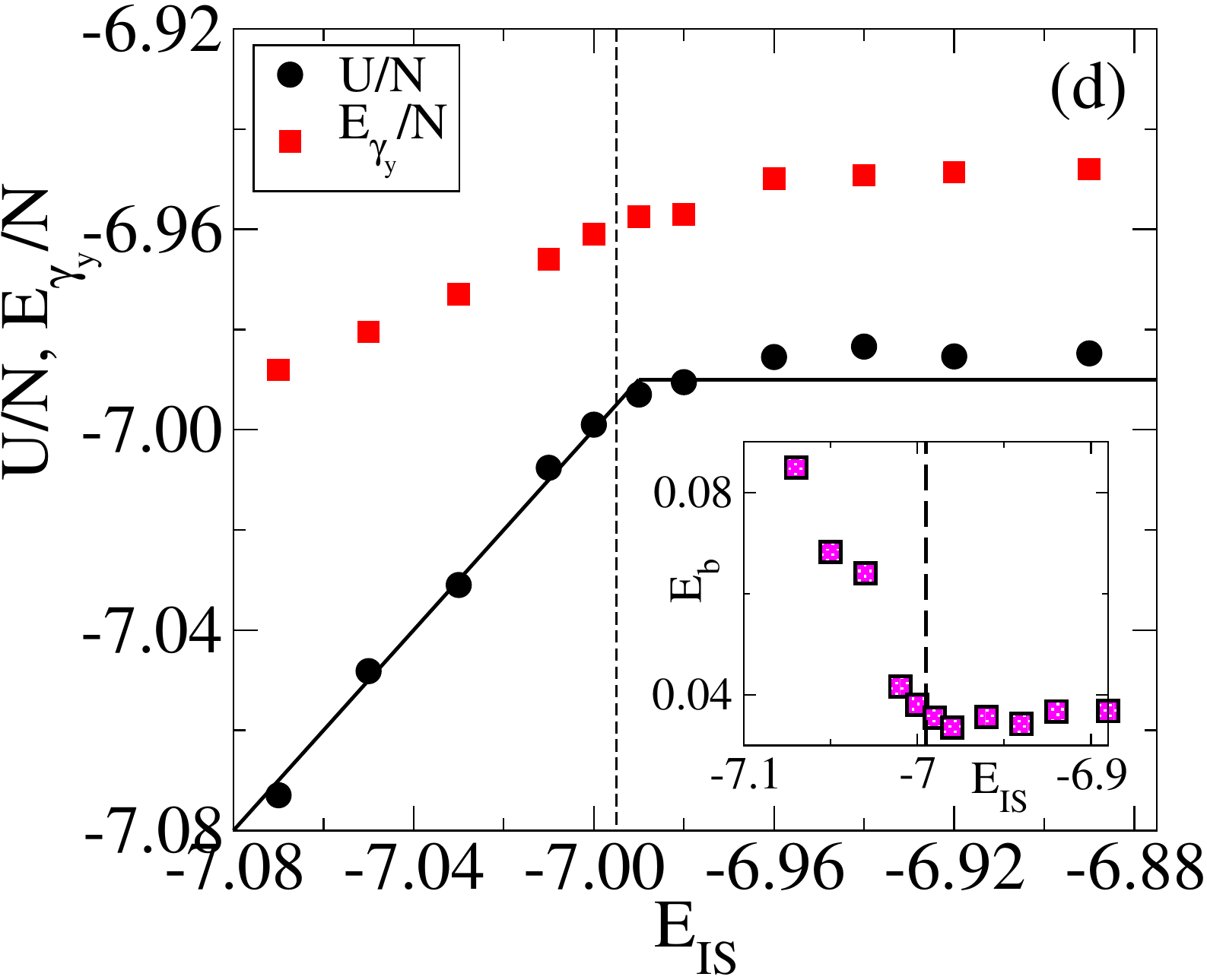}
  }
\caption{\label{fig_bmlj} {\bf Kob-Andersen BMLJ under cyclic shear:}
  ($N=4000$) (a) The steady state energies (stroboscopic) are plotted
  against $\gamma_{max}$ for different $E_{IS}$. (b) The maximum
  stress value is plotted against strain amplitude for different
  temperatures. Inset shows the jump in the energy, $\Delta E$, and
  the jump in maximum strain $\Delta\sigma$ at the yield strain, as a
  function of $E_{IS}$. (c) Plot of potential energy versus $\gamma$
  in the steady state, for different $E_{IS}$, at the critical strain
  amplitude. (d) Plot of energies $E_{\gamma_{y}}/N$ at $\gamma =
  \gamma_{max} = \gamma_{y}$, and $U/N$ (at $\gamma=0)$ for the yield
  amplitude $\gamma_{max} = \gamma_{y}$ as a function of
  $E_{IS}$. Inset shows the difference between $E_{\gamma_{y}}/N$ and
  $U/N$ {\it vs.} $E_{IS}$. Vertical lines in the insets are at
  $E_{IS} = -6.995$.}
\end{figure*}

Given that the cyclic shear results for silica appear qualitatively
different from the earlier results
\cite{leishangthem2017,parmarPRX2019} for the KA BMLJ, the natural
question to address is whether the earlier observations change when
one considers a broader range of IS energies for the initial
glasses. We address this by considering initial energies from $-6.89$
to $-7.07$, for $N = 4000$. The results shown in Fig. \ref{fig_bmlj}
reveal that indeed, by considering a larger range of annealing, the
picture changes radically from that observed earlier. As seen in
Fig. \ref{fig_bmlj} (a) the energies {\it vs.} strain amplitude for
different initial IS energies show a striking resemblance to the
behavior seen in BKS silica. For initial IS energies above $E_{IS} =
-6.99$, as the yield strain amplitude $ \gamma_{max} = \gamma_{y}$ is
approached, the energy of all the glasses converge to a common energy
value of $-6.995$, corresponding closely to a temperature of $T_{th} =
0.435$. For lower initial IS energies, there is a near lack of
annealing with strain amplitude below yielding, and a discontinuous
change in energies is observed at yielding. More striking in this case
compared to silica, however, is the significant increase in the yield
strain values with greater annealing, reaching a maximum value of
about $\gamma_{y} = 0.11$, compared to around $\gamma_{y} = 0.075$ for
the poorest annealed glasses. The difference in the two cases is
related to the more modest range of annealing achieved in the case of
silica (Additional analysis data pertaining to Fig.  \ref{fig_bmlj} is
found in Fig.s A5-A7 of {\it  Appendix}). Fig. \ref{fig_bmlj} (b)
shows that the jump in the maximum stress grows strongly with
annealing, and the inset of Fig. \ref{fig_bmlj} (b) shows that the
energy and stress jumps increase below $E_{IS} = -6.99$. In
Fig. \ref{fig_bmlj} (c) we show the steady state energies through the
cycle just below the respective yield strains
$\gamma_{y}$. Fig. \ref{fig_bmlj} (d) shows the energies at the
minimum, and at $\gamma_{y}$, and (in the inset) their difference,
revealing a striking change of behavior below the threshold energy of
$-6.995$.

Taken together, the results for silica and the KA BMLJ strongly
support a scenario that the nature of yielding changes drastically
below a threshold degree of annealing.  The main difference in
comparing with uniform shear is that under cyclic shear, poorly
annealed glasses evolve towards glasses at the thereshold degree of
annealing, and manifest a correspondingly mild discontinuous behavior.

\subsection*{Uniform shear  of the Kob-Andersen Binary Mixture Lennard-Jones Model}
\begin{figure*}[t]
\centerline{
  \includegraphics[width=.32\linewidth]{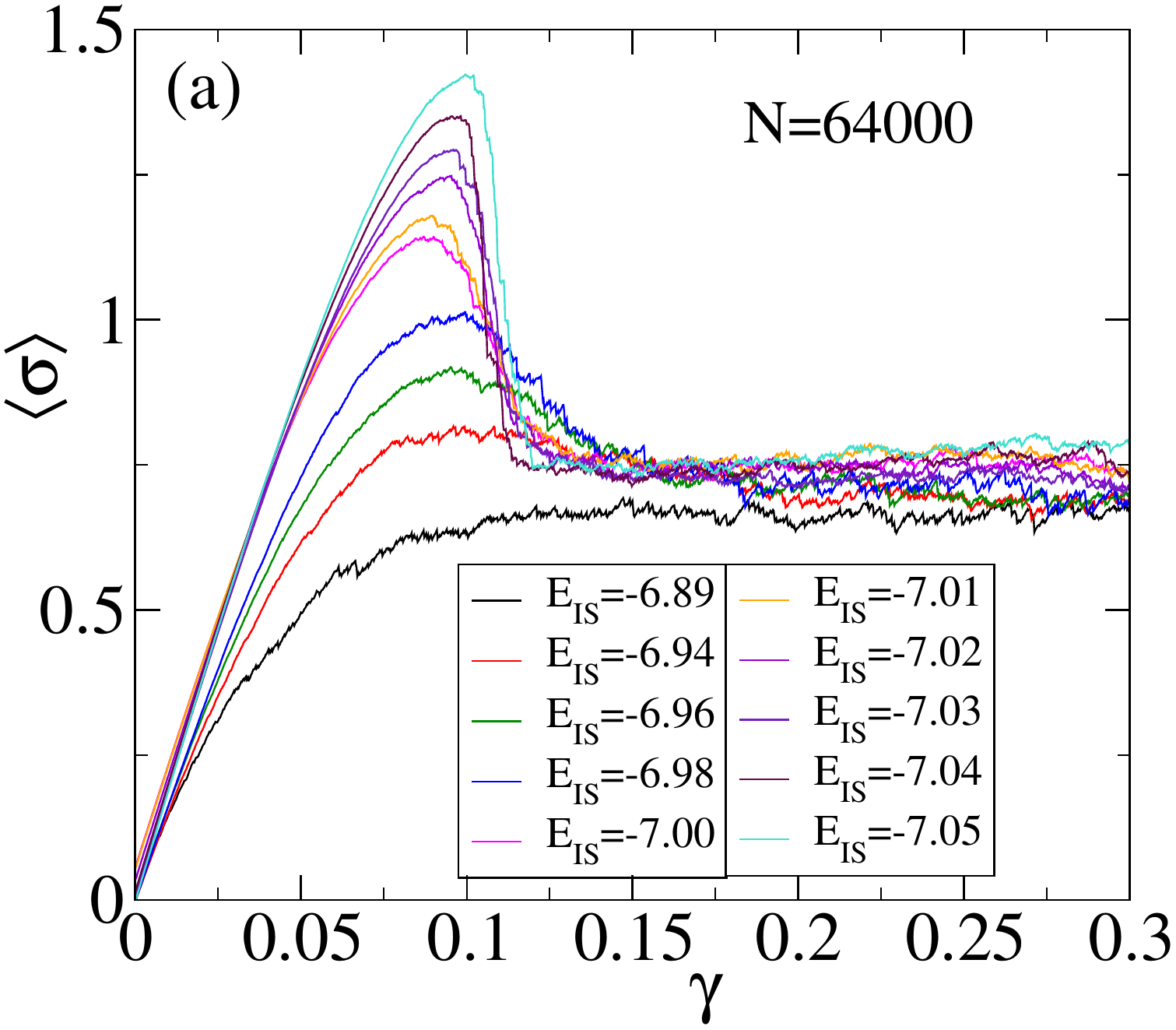}
  \includegraphics[width=.32\linewidth]{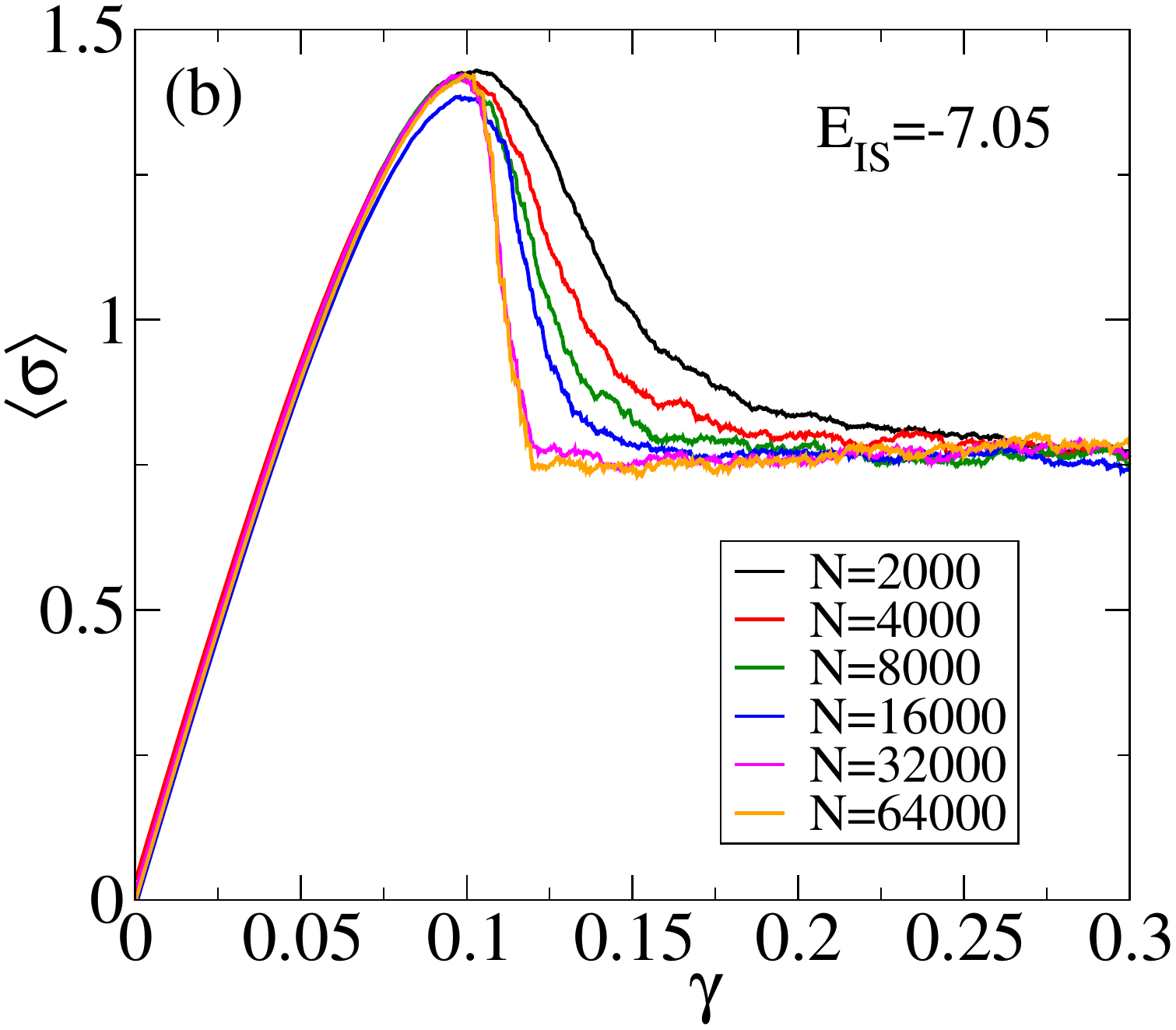}
  \includegraphics[width=.34\linewidth]{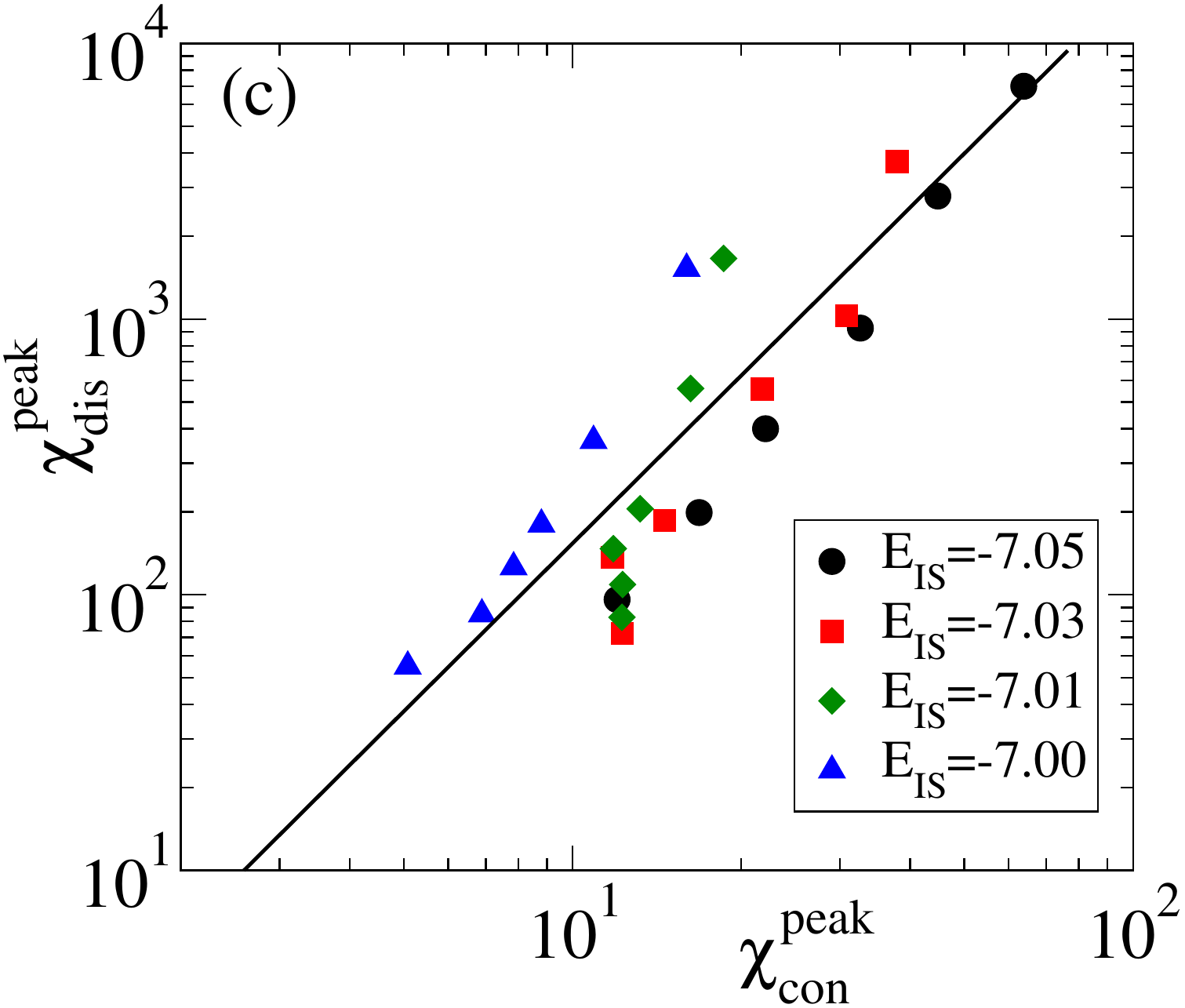}
  } 
  \caption{\label{fig_USbmlj} {\bf Kob-Andersen BMLJ under uniform
      shear:} (a) Stress-strain curves for different $E_{IS}$ for
    N=64000 averaged over several samples. (b) Stress-strain curves
    for $E_{IS} = -7.05$ for a range of system sizes.  (c) Scatter
    plot of the $\chi_{dis,peak}$ against $\chi_{con,peak}$ for
    different $E_{IS}$ below annealing threshold to show agreement
    with $\chi_{dis,peak} \sim \chi_{con,peak}^2$.}
\end{figure*}

Although results from cyclic shear simulations above provide results
that are consistent with the ductile to brittle transition scenario
discussed by Ozawa {\it et al.} \cite{ozawaPNAS2018}, neither model
has been studied in sufficient detail with uniform shear to allow a
direct comparison. In order to make such a comparison, we study the
response to uniform shear of the KA BMLJ model, for a wide range of
annealing of the glasses and system sizes, and obtain results
consistent with \cite{ozawaPNAS2018}. In Fig. \ref{fig_USbmlj} (a) we
show the sample averaged stress-strain curves for $N = 64000$ over the
studied range of glass energies. We see that as $E_{IS}$ decreases,
the yielding transition becomes progressively sharp.  Likewise, the
transition becomes sharper with system size, as shown in
Fig. \ref{fig_USbmlj} (b) for $E_{IS} = -7.05$, supporting a
discontinuous, first order transition in the thermodynamic limit. We
compute two associated susceptibilities, the "connected"
susceptibility $\chi_{con} = -{d \langle\sigma\rangle \over d \gamma}$
and the "disconnected" susceptibility $\chi_{dis} = N
[\langle\sigma^2\rangle - \langle\sigma\rangle^2]$, where the averages
are over samples (shown in {\it Appendix}, Fig. A8, along with
related results).  Both quantities grow with system size (with peak
values $\chi^{peak}_{dis} \sim N$, and $\chi^{peak}_{con} \sim
\sqrt{N}$), and as shown in Fig. \ref{fig_USbmlj} (c) are consistent
with $\chi^{peak}_{dis} \sim {\chi^{peak}_{con}}^{2}$, as would be
expected when disorder-induced sample to sample fluctuations dominate
in determining the susceptibilities \cite{ozawaPNAS2018}.

\subsection*{Dynamical Crossover}
\begin{figure*}[!t]
\centerline{
     \includegraphics[width=.39\linewidth]{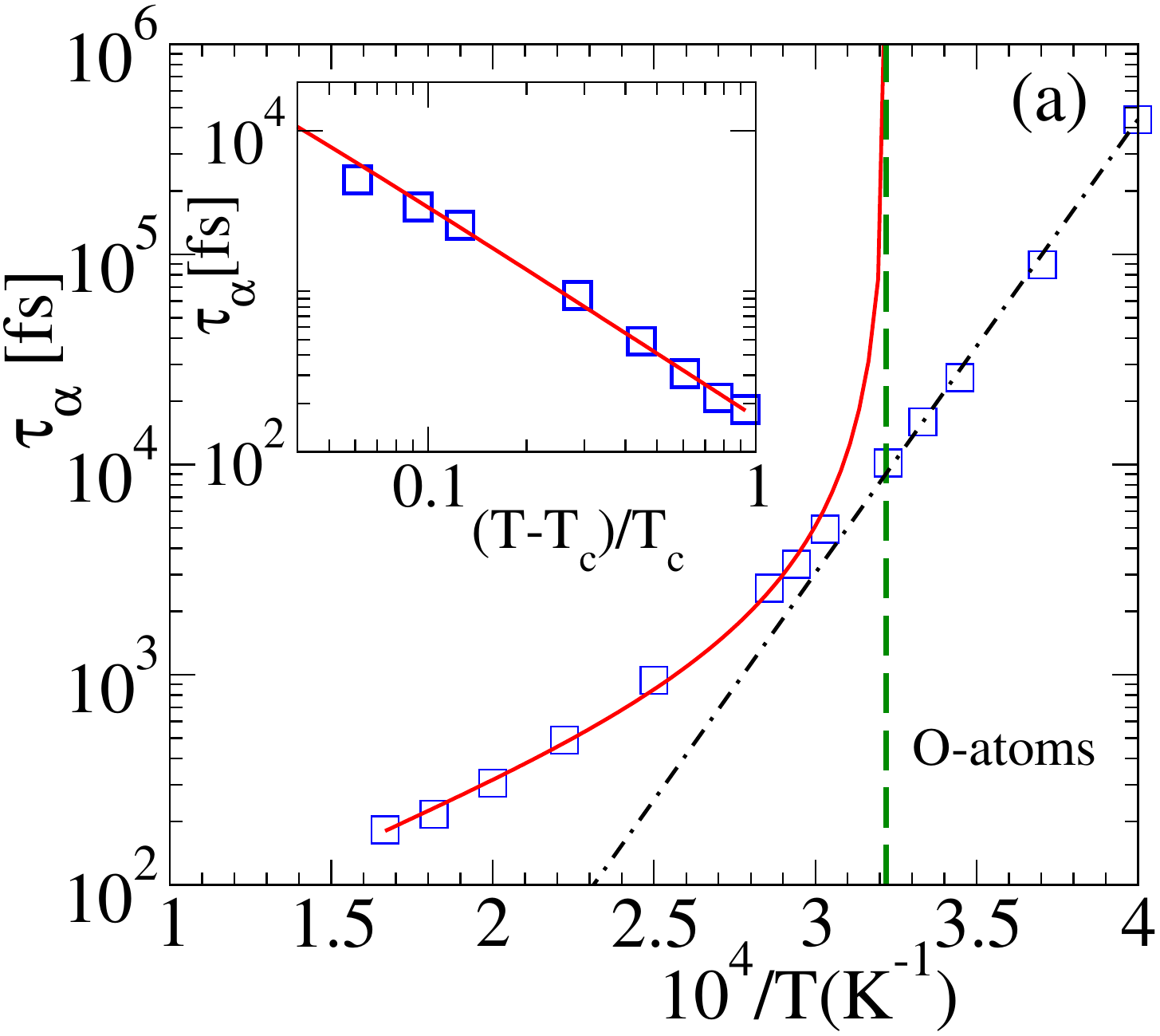}
     \includegraphics[width=.39\linewidth]{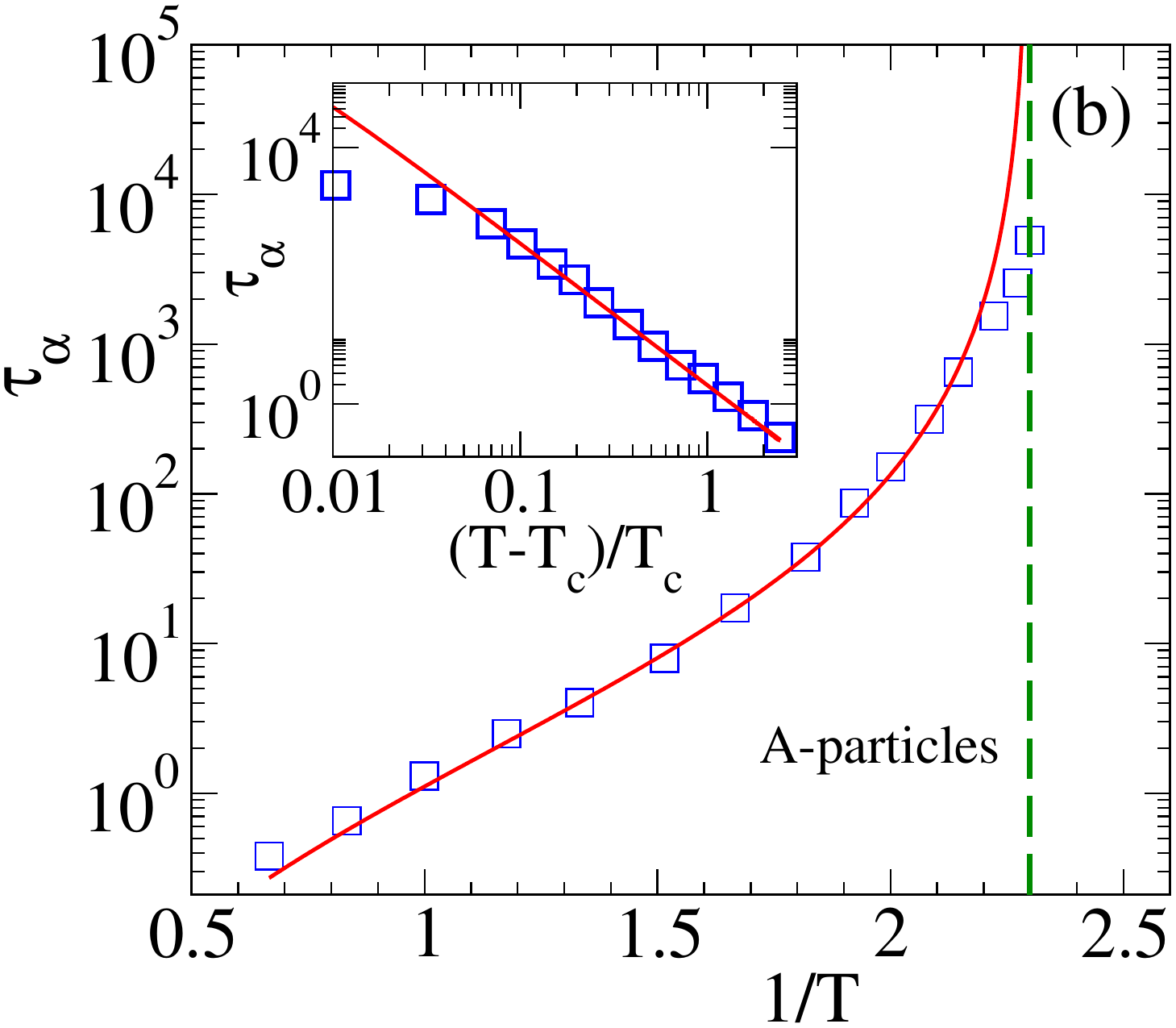}
     }
   \caption{\label{fig_dyncross} {\bf Dynamical crossover}: Plots of
     $\tau_{\alpha}$ extracted from $F_S(k,t)$, shown as a function of
     the inverse temperature on a semi-logarithmic scale, for (a)
     silica and (b) BMLJ. The red solid lines through the data points
     are fits of the form $\tau_{\alpha} = \tau_0
     (T-T_c)^{-\gamma}$. For silica, the value of $T_c$ and $\gamma$
     are $3100 K$ (shown by the vertical dashed line) and $1.5$ for
     the O-atom, and the dotted dashed line represents the Arrhenius
     temperature dependence low the temperature regime. For BMLJ,
     $T_c$ and $\gamma$ are found to be $0.435$ and $2.26$,
     respectively, for the A-particle. }
\end{figure*}
The threshold annealing level for BKS silica and the KA BMLJ
correspond respectively to $T = 3100 K$, and $T = 0.435$. To
investigate the meaning of these crossover temperatures we consider
the dynamics of the respective liquids. In Fig. \ref{fig_dyncross}, we
show the $\alpha$ relaxation times of the two systems {\it vs.}
temperature. The self part of the intermediate scattering functions
from which the relaxation times are obtained are shown in Fig. A9 in
{\it Appendix}.  For silica, we see that the relaxation times
$\tau_{\alpha}$ show a fragile to strong crossover, as reported in
several works previously
\cite{horbachPRB99,voivodNAT01,voivodPRE04a,heuerPRL04}, at $ T = 3100
K$, below which their temperature dependence becomes
Arrhenius. Relaxation times above this cross over temperature are well
described by a power law form $\tau_{\alpha} \sim (T -
T_c)^{-\gamma}$, as previously discussed
\cite{horbachPRB99,horbachPRE01,voigtmannJPCM08,horbachJPCM08}, with
$T_c = 3100 K$, and $\gamma = 1.5$. The value of $\gamma$ is lower
than reported in previous mode coupling analyses, which we do not
attempt here. Instead, we simply note this as evidence for a fragile
to strong crossover. In the case of KA BMLJ, previous analyses have
estimated the mode coupling crossover temperature to be $T_c \sim
0.435$, and a fit to our data supports this identification. A $\gamma$
value of $2.26$ best describes our data, which is similar to values
reported from mode coupling analyses previously
\cite{kob1994,kob1995,szamel2005}.

Thus, in both cases, we conclude that the threshold degree of
annealing corresponds to a crossover temperature in the dynamics in
the corresponding liquids. Although such an identification is merely
an observation at present, it is made plausible and meaningful by
numerous earlier studies, with associate the dynamical crossovers with
a transition from diffusive to activated dynamics, and
correspondingly, a change in the nature of the energy landscape
sampled
\cite{tbsJCP2000,broderixPRL00,leonardoPRL,angelanioiJCP02,LanaveSilica2002,heuerPRL04}. Our
results suggest that it may be fruitful to seek the common causes of
the transition to brittle behavior of athermal glasses and changes in
the character of dynamics in the corresponding liquids\cite{Yeh2019}.

\subsection*{Summary} 

In summary, we have analysed the nature of yielding under cyclic shear
deformation of two model glasses, the BKS silica glass, and the atomic
model KA BMLJ, and found that for sufficiently well annealed glasses,
yielding is strongly discontinuous, with the magnitude of
discontinuity growing with the degree of annealing. Cyclic shear
further offers a way for precisely identifying the threshold
annealing/disorder through the qualitatively distinct responses above
and below the annealing threshold. We show that for one of the models
studied (KA BMLJ), uniform shear deformation reveals a change from
ductile to brittle crossover that is consistent both with earlier
analysis \cite{ozawaPNAS2018} and with the results of cyclic shear
deformation. Finally, we find that the annealing threshold corresponds
to well known dynamical crossover temperatures in the two systems we
study, an observation that merits further analysis to understand their
possible common origins. Whether the present results enable analysis
of yielding at finite temperatures and strain rates, at least for well
annealed glasses, along lines explored for crystals
\cite{sausset2010solids,nath2018existence} is another interesting question
to pursue.

\appendix

\section*{APPENDIX}
\subsection{Models}
The BKS potential \cite{beestPRL90} of silica we study, with
modifications\cite{voivodPRE04a} to avoid unwanted divergences and
treat long range forces properly, is given by
\begin{eqnarray}
\label{mbks}
U(r_{ij})&=&\frac{1}{4\pi\epsilon_0}\frac{q_iq_j}{r_{ij}}\nonumber\\&+&\left \{ \begin{array}{lll}
A_{ij}\exp(-B_{ij}r_{ij})-C_{ij}r_{ij}^{-6}+ \phi(r_{ij})&r_{ij}\le
r_s\\ \sum_{k=3}^{5}D_{ij}^k(r_{ij}-R_c)^k&r_s<r_{ij}<r_c\\ 0&r_{ij}\ge
r_c\end{array} \right.
\end{eqnarray}
where, the short-range term $\phi(r_{ij})=4\epsilon_{ij}\left[ \left(
  \frac{\sigma_{ij}}{r_{ij}}\right)^{30}-
  \left(\frac{\sigma_{ij}}{r_{ij}}\right)^{6}\right]$ is added for
$r<r_s$ to prevent a negative divergence. The Coulomb term is
evaluated by the Ewald summation technique with Ewald parameter
$\alpha=2.5 \AA^{-1}$. The momentum space summation is carried out to
a radius of six reciprocal lattice cell widths. The real part of the
potential is truncated at $r_s=7.7747\AA$. In the $r_s<r<r_c$ regime
with $r_c=10\AA$, a fifth order polynomial is added to make the
potential go to zero smoothly. The coefficients of $\phi(r_{ij})$ are
chosen such that the modified potential has no inflection at small
$r$. The values of all the co-efficient can be found in
Ref. \cite{voivodPRE04a}. An integration time step of $1 fs$ is used.

The Kob-Andersen binary (80:20) mixture of Lennard Jones particles we
study, with the interaction potential truncated at a cutoff distance
of $r_{c \alpha\beta}=2.5 \sigma_{\alpha \beta}$ such that both the
potential and the force smoothly go to zero, is given by
\begin{eqnarray}
V_{\alpha\beta}(r)&=&4 \epsilon_{\alpha\beta} \left[ \left(
  \frac{\sigma_{\alpha\beta}}{r} \right)^{12} - \left(
  \frac{\sigma_{\alpha\beta}}{r} \right)^{6} \right]\\ \nonumber & +& 4
\epsilon_{\alpha\beta}\left[c_{0 \alpha\beta} + c_{2
    \alpha\beta}\left(\frac{r}{\sigma_{\alpha\beta}}\right)^{2}\right],
r_{\alpha\beta} < r_{c\alpha\beta}
\label{eqn:KABMLJmodel}
\end{eqnarray}
where $\alpha,\beta \in \{A,B\}$ and the parameters
$\epsilon_{AB}/\epsilon_{AA}=1.5$, $\epsilon_{BB}/\epsilon_{AA}=0.5$,
$\sigma_{AB}/\sigma_{AA}=0.80$, $\sigma_{BB}/\sigma_{AA}=0.88$.
Energy and length are in the units of $\epsilon_{AA}$ and
$\sigma_{AA}$ respectively, and likewise, reduced units are used for
other quantities. An integration time step of $0.005$ is used.

\subsection{Initial glass preparation}
For both the models we performed constant volume, temperature (NVT)
molecular dynamics simulations using the Nose-Hoover thermostat. The
systems are equilibrated for $20\tau_{\alpha}$ for low temperatures,
and for $10^3\tau_{\alpha}$, for the high temperature for silica, and
more than $100 \tau_{\alpha}$ for the KA BMLJ, $\tau_{\alpha}$ being
the structural relaxation time (see Fig. A9, {\it Appendix}).

For the KA BMLJ system, we generate inherent structures at energies
lower than $-7.00$ through the finite temperature, shear rate
annealing protocol explored in \cite{das2018annealing}, with shear
rate $\dot{\gamma} = 10^{-5}$, strain amplitude $\gamma_{max} = 0.035$
and simulation temperature $T=0.3$. Further details of this procedure
may be found in \cite{das2018annealing}.

Independent configurations (between 5 and 12 for silica, and 6 for KA
BMLJ) sampled from the equilibrium trajectory are subjected to energy
minimization to obtain sets of inherent structures (IS) to be used for
cyclic shear. Between $24$ and $200$ for $E_{IS} \leq -7.00$ and $12$
for $E_{IS} > -7.00$ are used for uniform shear simulations. Energy
minimization is always (including AQS) performed using the
conjugate-gradient algorithm.

\bibliographystyle{apsrev4-1} 
\bibliography{ref}
\clearpage

\onecolumngrid
\section*{Appendix}
\renewcommand{\thefigure}{A\arabic{figure}}
\setcounter{subsection}{0} 
\setcounter{figure}{0} 
\hrule
\vspace{.5cm}

\subsection{Inherent structure energy evolution with temperature}
\begin{figure*}[h]
  \centerline{
    \includegraphics[width=.44\linewidth]{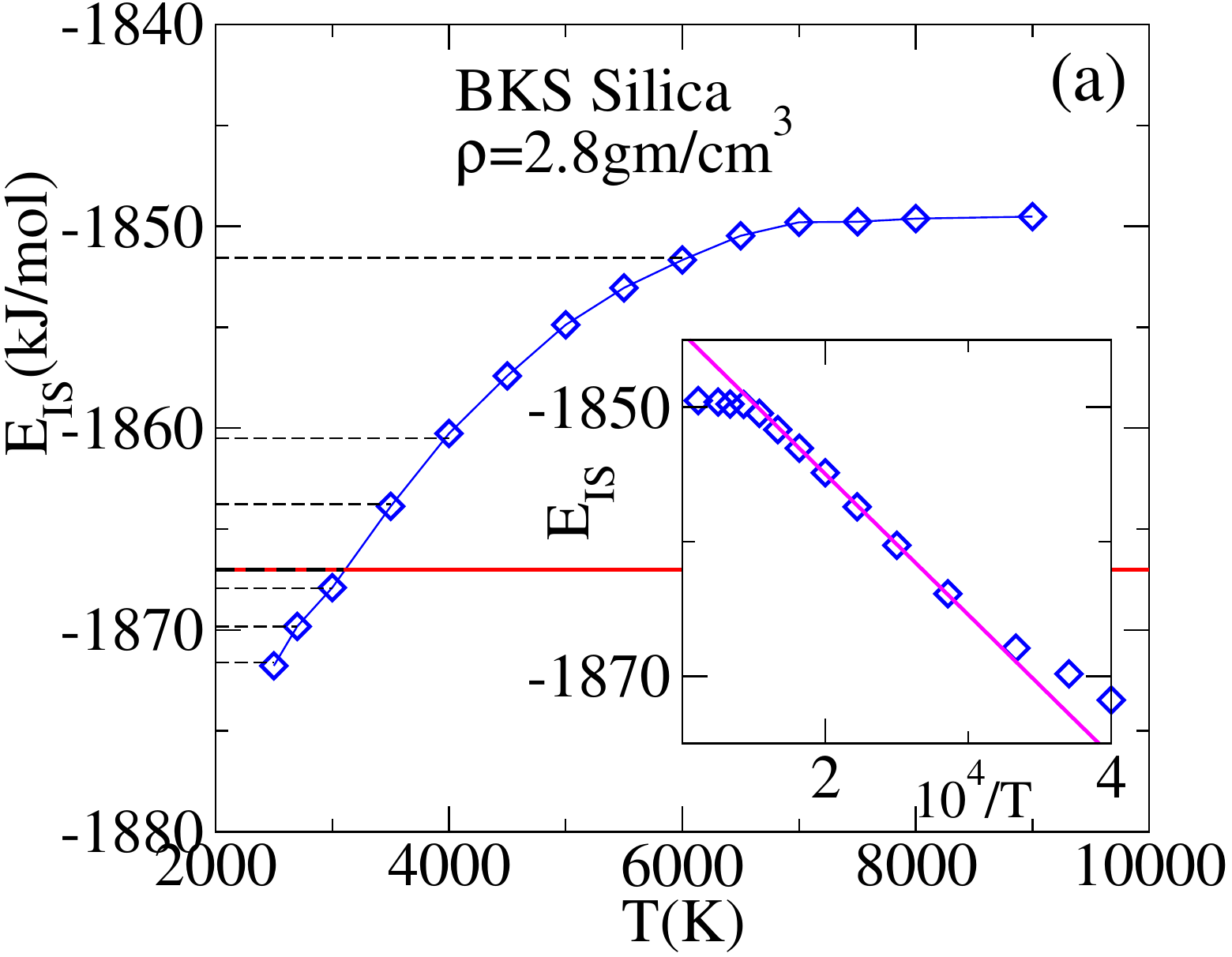}
    \includegraphics[width=.4\linewidth]{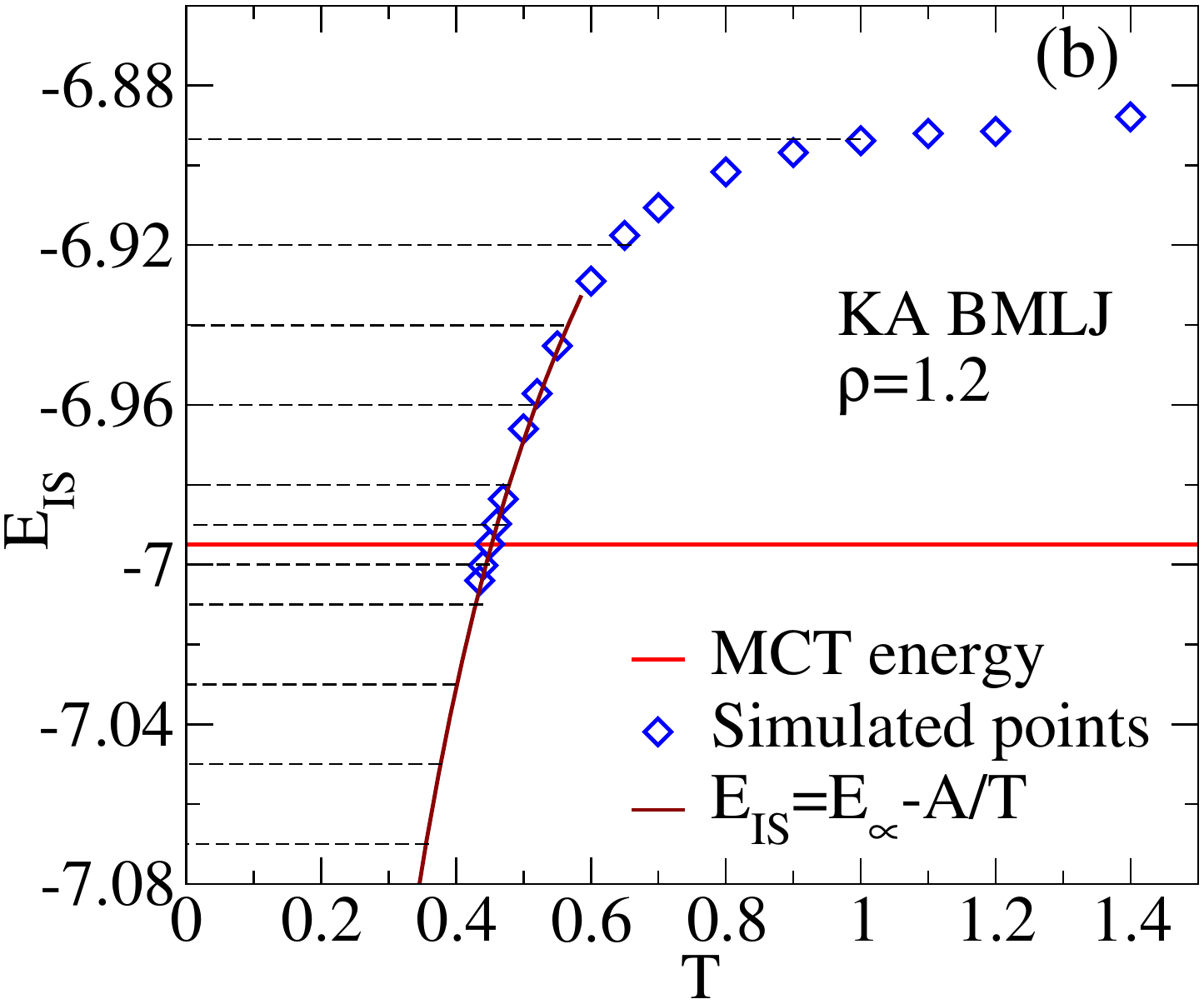}
    }
    \caption{\label{SI_EIS_T_silica} Plot of inherent structure energy
      against temperature $T$ for (a) silica and (b) BMLJ. For silica,
      each data point is sampled over $400$ minimized
      configurations. For KA BMLJ the extrapolated curve is obtained
      by fitting equilibrium molecular dynamics simulation data to the
      equation: $E_{\infty}-A/T$ below the temperature $T = 0.7$ to
      obtain a mapping between the IS energy and temperature in the
      low temperature regime. Dashed horizontal lines represent the
      energy levels from which the IS-configurations are sampled to
      perform cyclic and uniform shear deformation. Horizontal red
      solid lines are indicating the threshold energy levels in the
      two models.}
\end{figure*}

Here we show the behaviour of inherent structure energy $E_{IS}$ for a
wide range of temperatures for Silica and BMLJ, in
Fig. \ref{SI_EIS_T_silica} (a) and (b). For sufficiently high
temperature $T$, the values of $E_{IS}$ are show a plateau. For
silica, $E_{IS}$ data shows deviations from $1/T$ behaviour at low $T$
as shown in the inset, discussed as an indication of the
fragile-to-strong crossover in Ref. \cite{voivodNAT01}.

\subsection{Energy as a function of accumulated strain for BKS Silica}
\begin{figure*}[h]
  \centerline{
        \includegraphics[width=.55\linewidth]{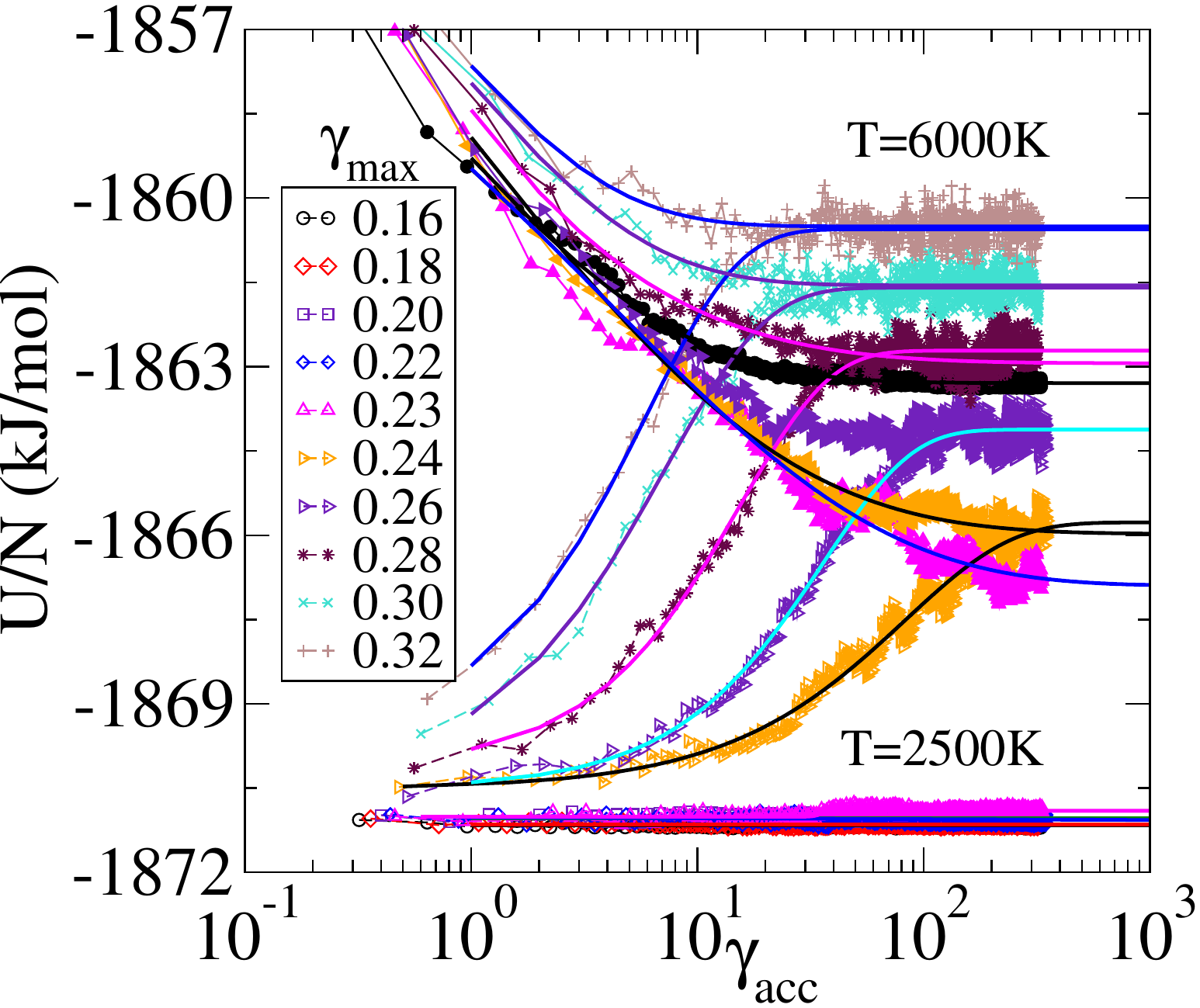}
     }
    \caption{\label{SI_sstate_silica} Potential energy per particle
      $U/N$ for zero-strain configurations, for $T=2500$K (lower set
      of curves, open symbols) and $T=6000$K (upper set of curves,
      filled symbols) for several strain amplitudes
      $\gamma_{max}$. Smooth solid lines through each data set are
      fits to a stretched exponential form from which the steady state
      energy values reported in Fig. 1 in the main text are
      obtained. }
\end{figure*}

The energy of the stroboscopic ($\gamma=0$) configurations are shown
in Fig. \ref{SI_sstate_silica}, as a function of the accumulated
strain. Data are presented for both $T=2500$K and $T=6000$K. As the
yielding amplitude $\gamma_{max}=\gamma_{y}=0.23$ is approached, the
number of cycles to attain the steady state increases strongly.  For
both the temperatures, when $\gamma_{max}>\gamma_{y}=0.23$, the same
final energies are reached irrespective of initial values, which is
not true for $\gamma_{max} < \gamma_{y}=0.23$.

\subsection{Stress-strain behavior over a cycle for BKS Silica}
\begin{figure*}[h]
  \centerline{
    \includegraphics[width=.4\linewidth]{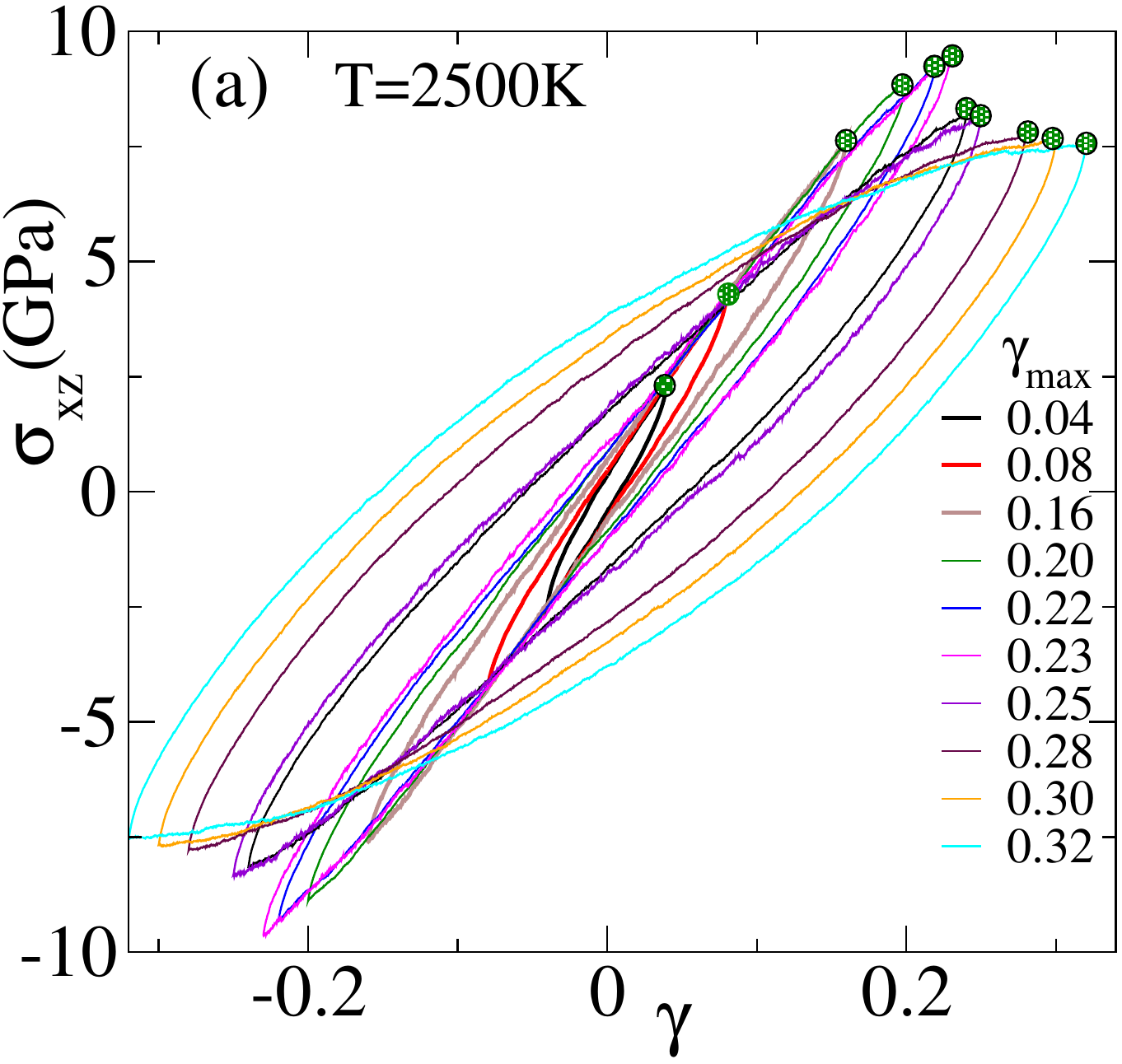}
     \includegraphics[width=.4\linewidth]{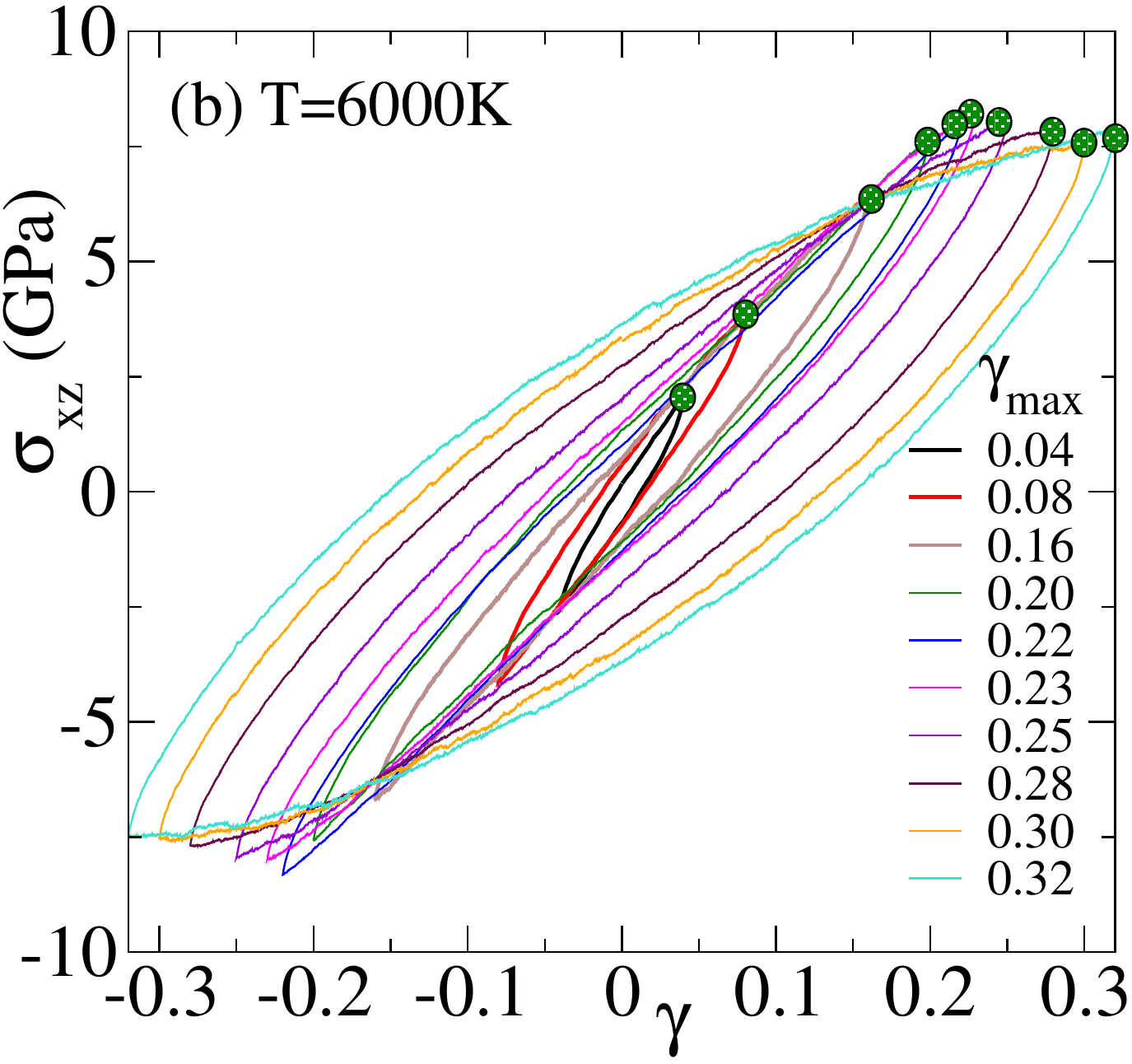}
    }
    \caption{\label{SI_stress_cycle_silica} {\bf Stress strain
        behaviour of Silica under cyclic shear:} Plot of averaged
      stress against $\gamma$ over a full cycle of deformation for
      different values of $\gamma_{\rm{max}}$ for (a) $T=2500$K and
      (b) $T=6000$K. The curves are averaged over $10$ samples and
      several cycles ($\approx 30$) for each sample in the steady
      state. Green dots represent the value of the maximum stress, at
      the strain amplitude, which are reported in Fig. 1 of the main
      text.}
\end{figure*}

The stress-strain curves over a full-cycle for different strain
amplitudes are shown in Fig. \ref{SI_stress_cycle_silica} for two
different temperatures. The average stress values are calculated in
the steady state.

\subsection{Potential energy as a function of  strain over multiple cycles for BKS silica}
\begin{figure*}[h]
  \centerline{
    \includegraphics[width=.45\linewidth]{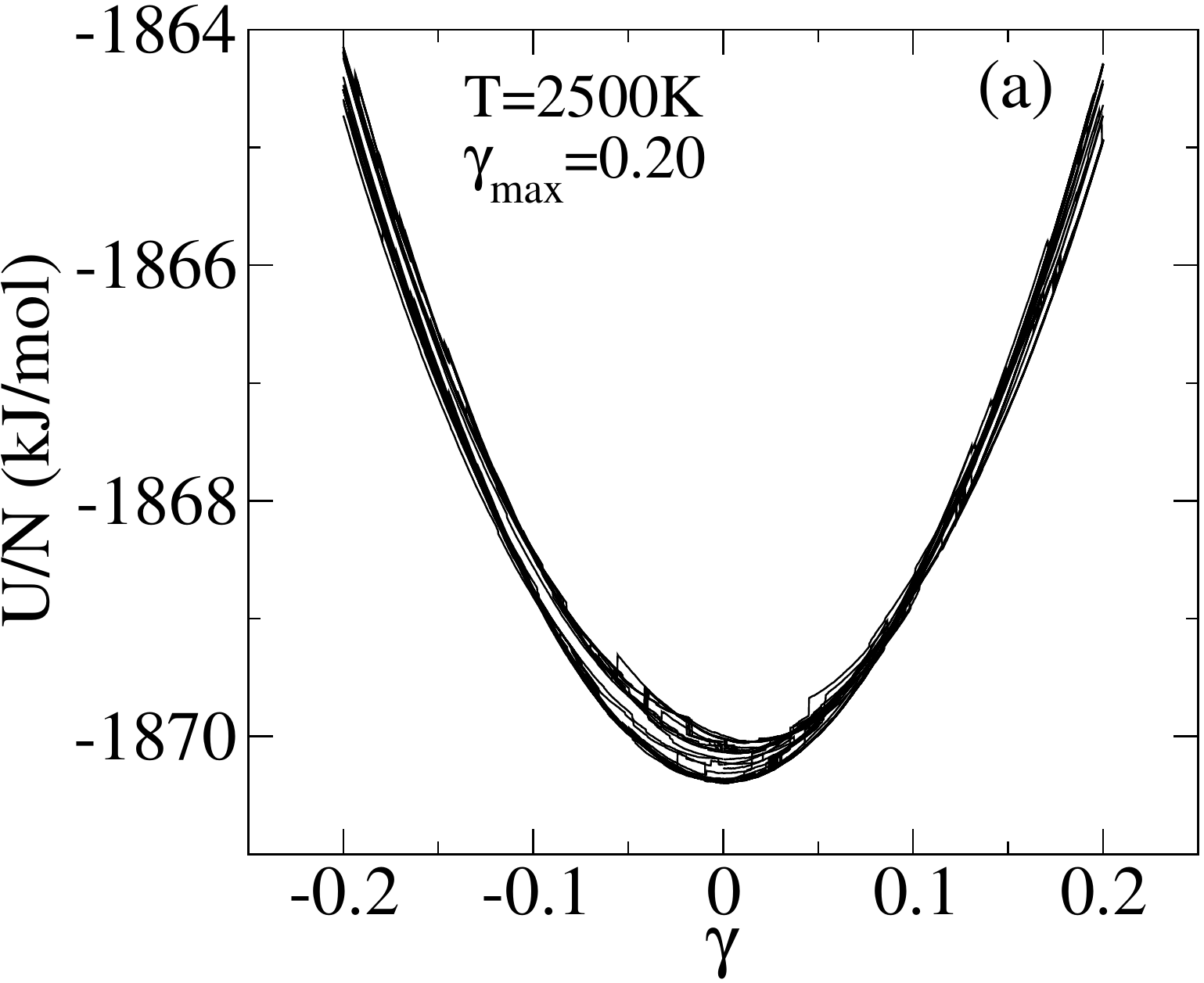}
    \includegraphics[width=.45\linewidth]{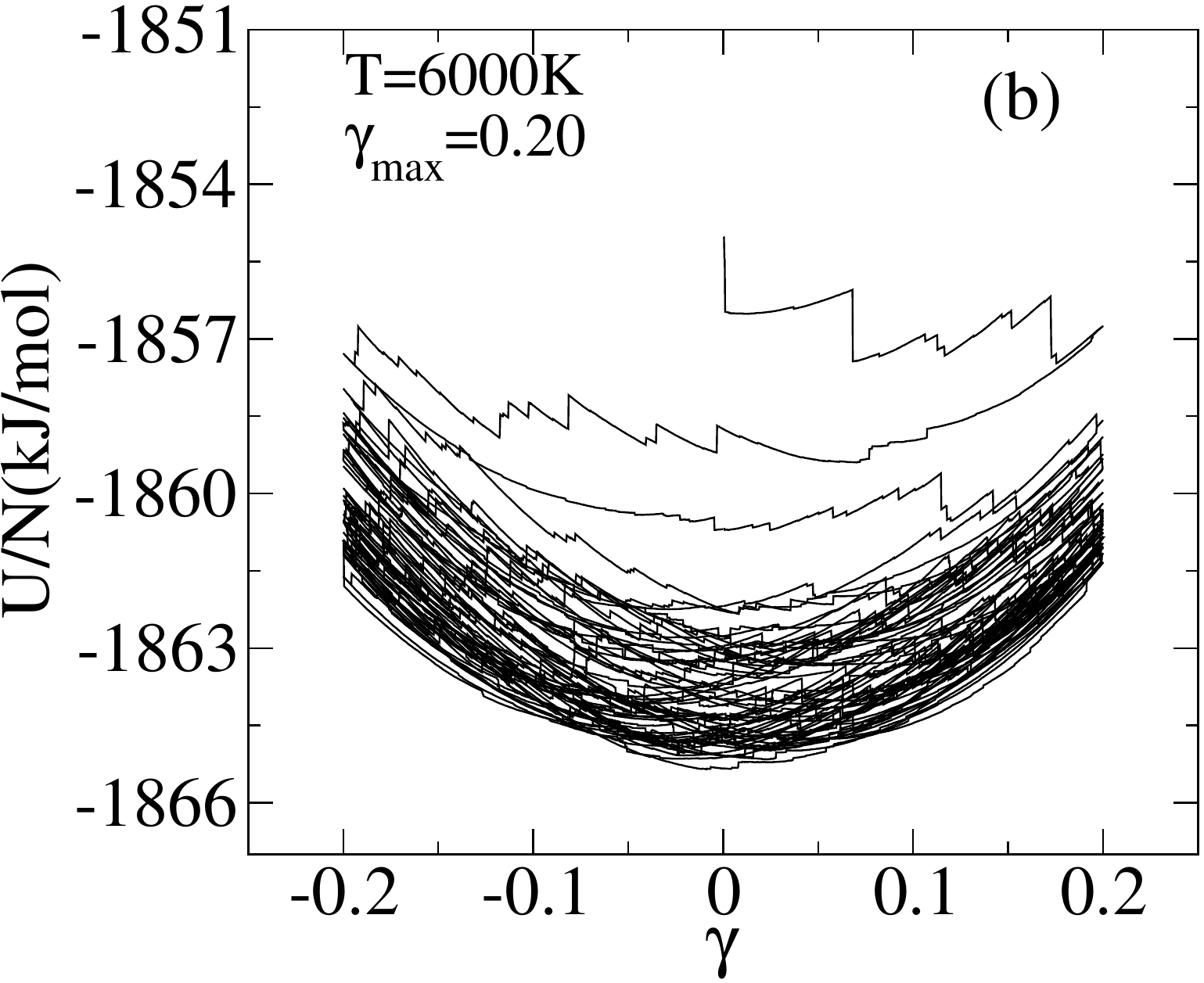}
    }
    \caption{\label{SI_PE_cycle_BKS}{\textbf{Potential energy {\it
            vs.} strain over multiple cycles for silica}:} Plot of the
      potential energy over the cycles from the beginning of the
      simulations for (a) $T=2500$ and (b) $T=6000K$, at
      $\gamma_{max}=0.20$.}
\end{figure*}

In Fig. \ref{SI_PE_cycle_BKS}, we show the evolution of the energy for
two temperatures for a strain amplitude $\gamma_{max}=0.20$ below
yielding. For the low temperature case ($T=2500$K) the system explores
almost the same energy path for all cycles. For high temperature
($T=6000$K), starting from a high energy state, the system anneals to
a lower energy value around which it eventually remains oscillating.

\subsection{Energy as a function of accumulated strain for the Kob-Andersen BMLJ}
\begin{figure*}[h]
\centerline{
  \includegraphics[width=.44\linewidth]{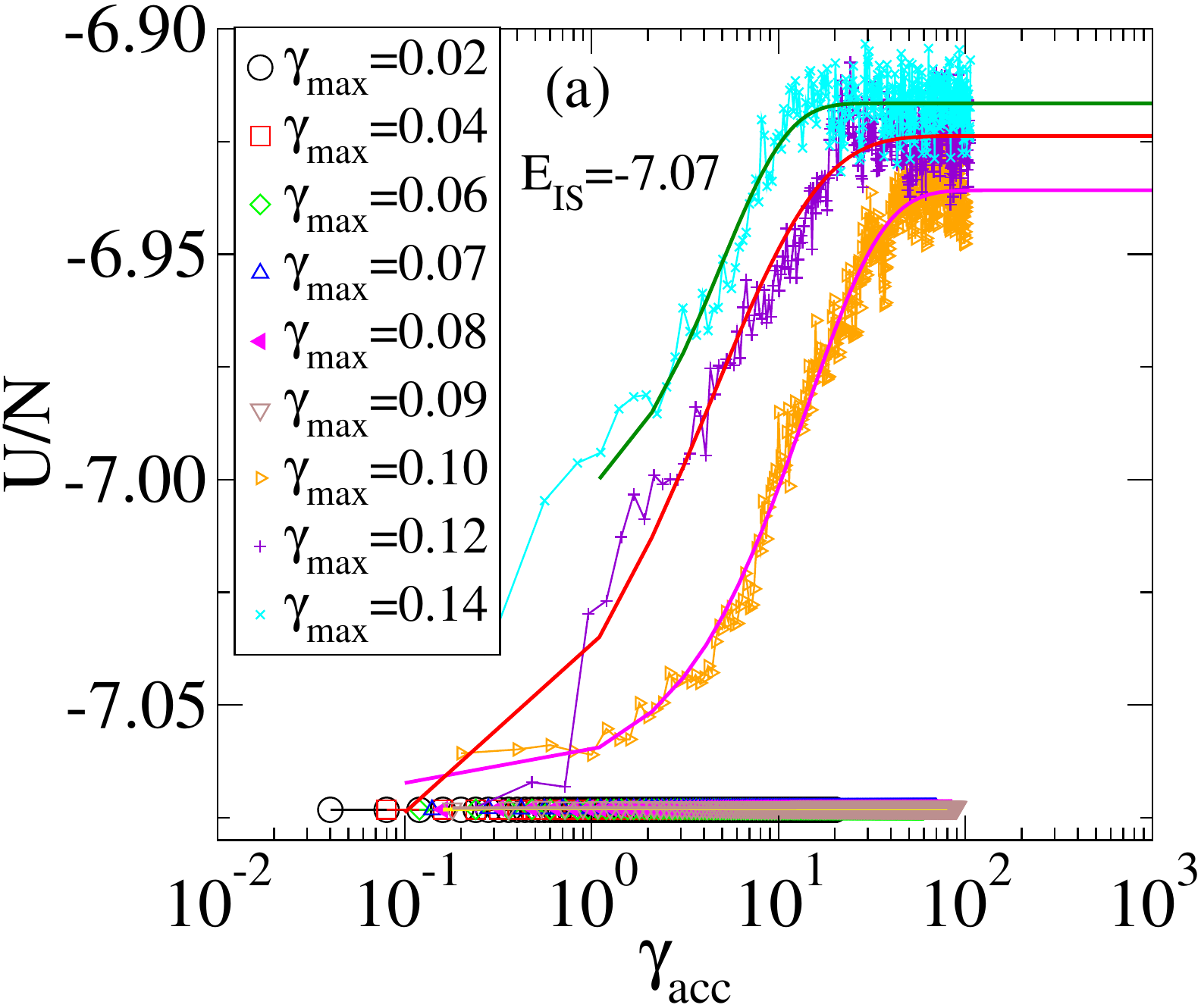}
     \includegraphics[width=.44\linewidth]{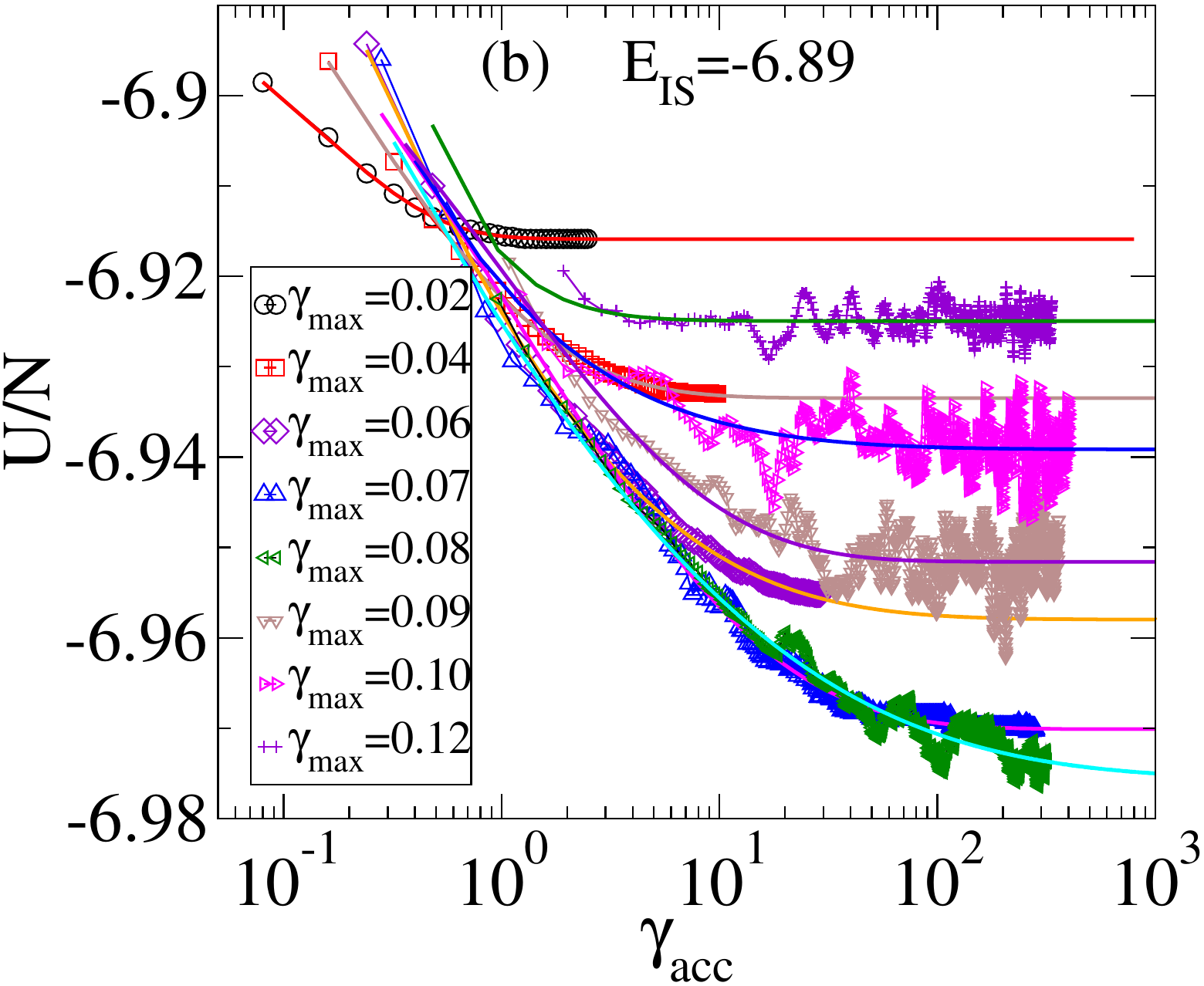}
    }
\caption{\label{SI_state_BMLJ} {\bf Approach to steady state for KA
    BMLJ:} Potential energy per particle $U/N$ for zero-strain
  configurations plotted against accumulated strain $\gamma_{acc}$,
  (a) for $E_{IS}=-7.07$ and (b)for $E_{IS}=-6.89$. The data are
  averaged over $6$ samples for the system size $N=4000$. Solid lines
  through each data set are fits to a stretched exponential form.}
\end{figure*}

In Fig. \ref{SI_state_BMLJ}, we show the potential energy per particle
$U/N$ for stroboscopic configurations, as a function of the
accumulated strain $\gamma_{acc}$ for two differently annealed BMLJ
systems with $E_{IS} = -7.07$ and $-6.89$. The steady state value of
$U/N$ is obtained by fitting the data to a stretched exponential form,
which are shown in Fig. 2 of the main text. For the well annealed
case, below yielding amplitude, the energy of the system remains
essentially constant. Above the $\gamma_y$, the energy increases with
the number of shear cycles and reaches a steady state at longer
times. For the poorly annealed glass, the energy always decreases with
shear deformation cycles until it reaches a steady state.

\subsection{Stress-strain behavior over a cycle for  Kob-Anderson BMLJ}
\begin{figure*}[h]
\centerline{
  \includegraphics[width=.44\linewidth]{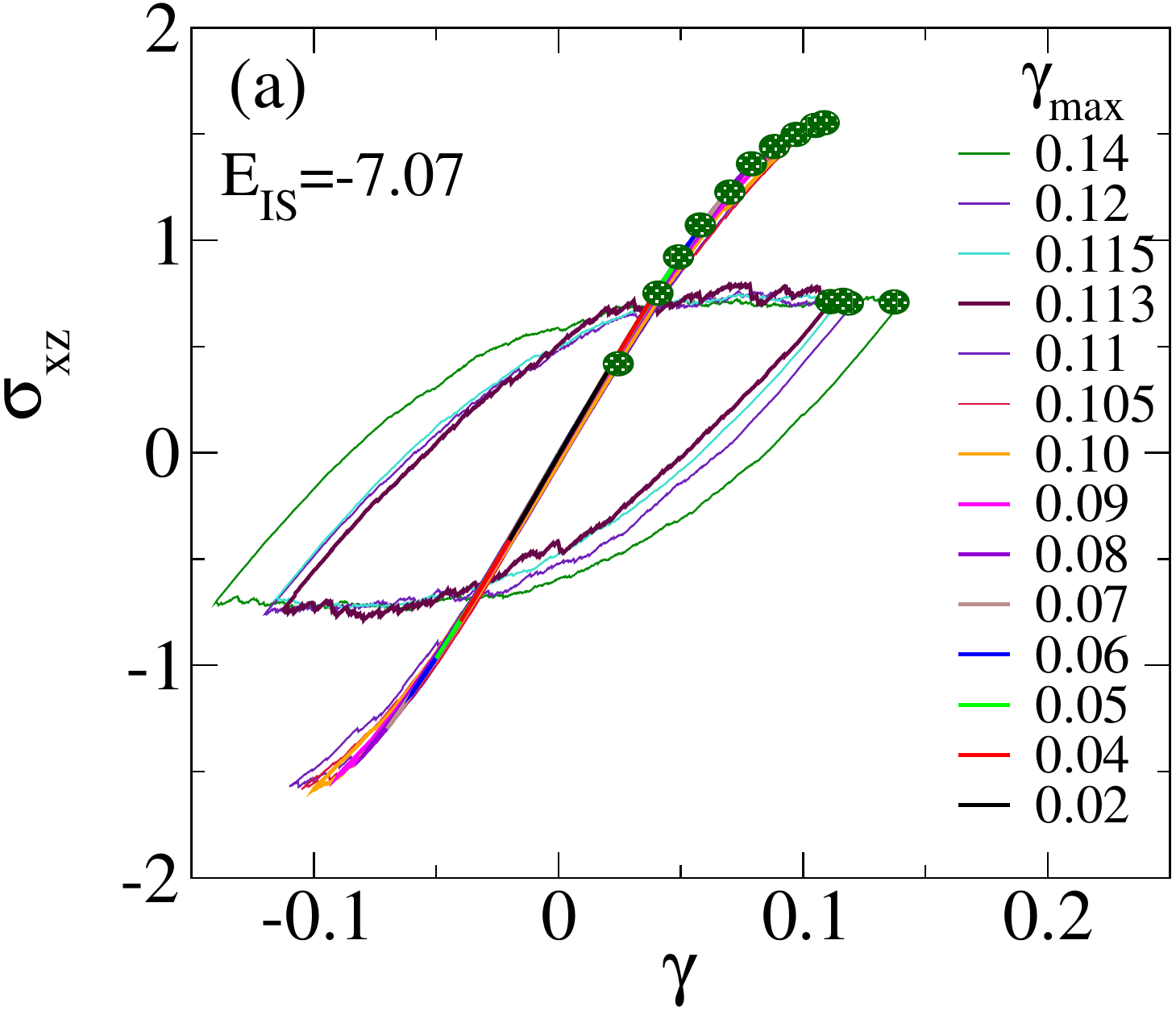}
    \includegraphics[width=.44\linewidth]{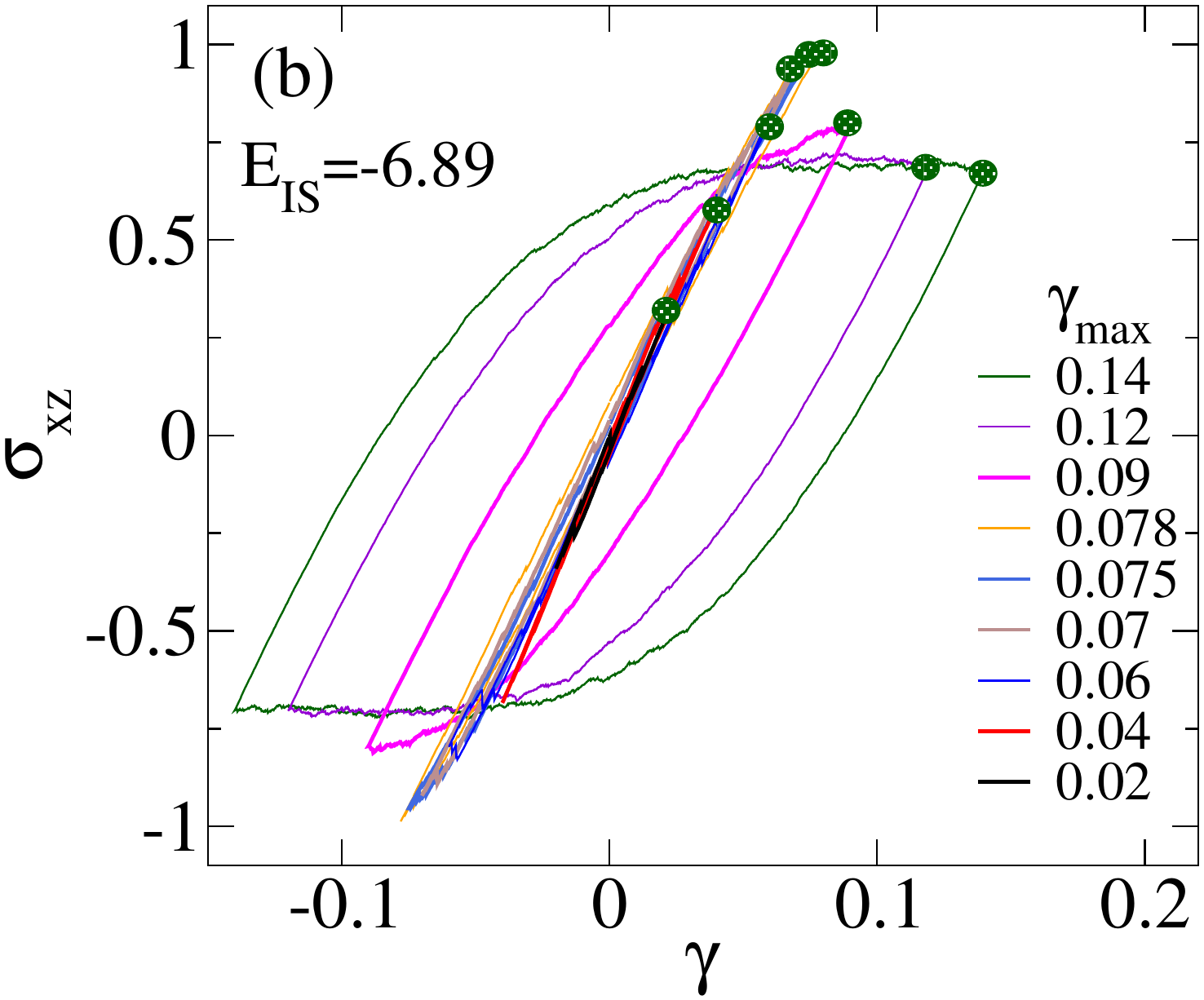}
    }
\caption{\label{SI_stress_cycle_BMLJ} {\bf Stress strain behavior of
    BMLJ under cyclic shear:} Plot of stress as a function of $\gamma$
  over a full cycle of deformation for different strain amplitudes
  $\gamma_{max}$, averaged over several cycles in the steady state for
  (a) $E_{IS}=-7.07$ and (b) $E_{IS}=-6.89$. Green dots represent the
  maximum of stress reached by the system at the strain amplitude
  $\gamma=\gamma_{max}$.}
\end{figure*}

The averaged stress-strain curves in the steady state for a full-cycle
of deformation for different strain amplitudes are shown in
Fig. \ref{SI_stress_cycle_BMLJ} for two different energies.

\subsection{Potential energy as a function of  strain over multiple cycles for the Kob-Andersen BMLJ}
\begin{figure*}[h]
     \centerline{
     \includegraphics[width=.4\linewidth]{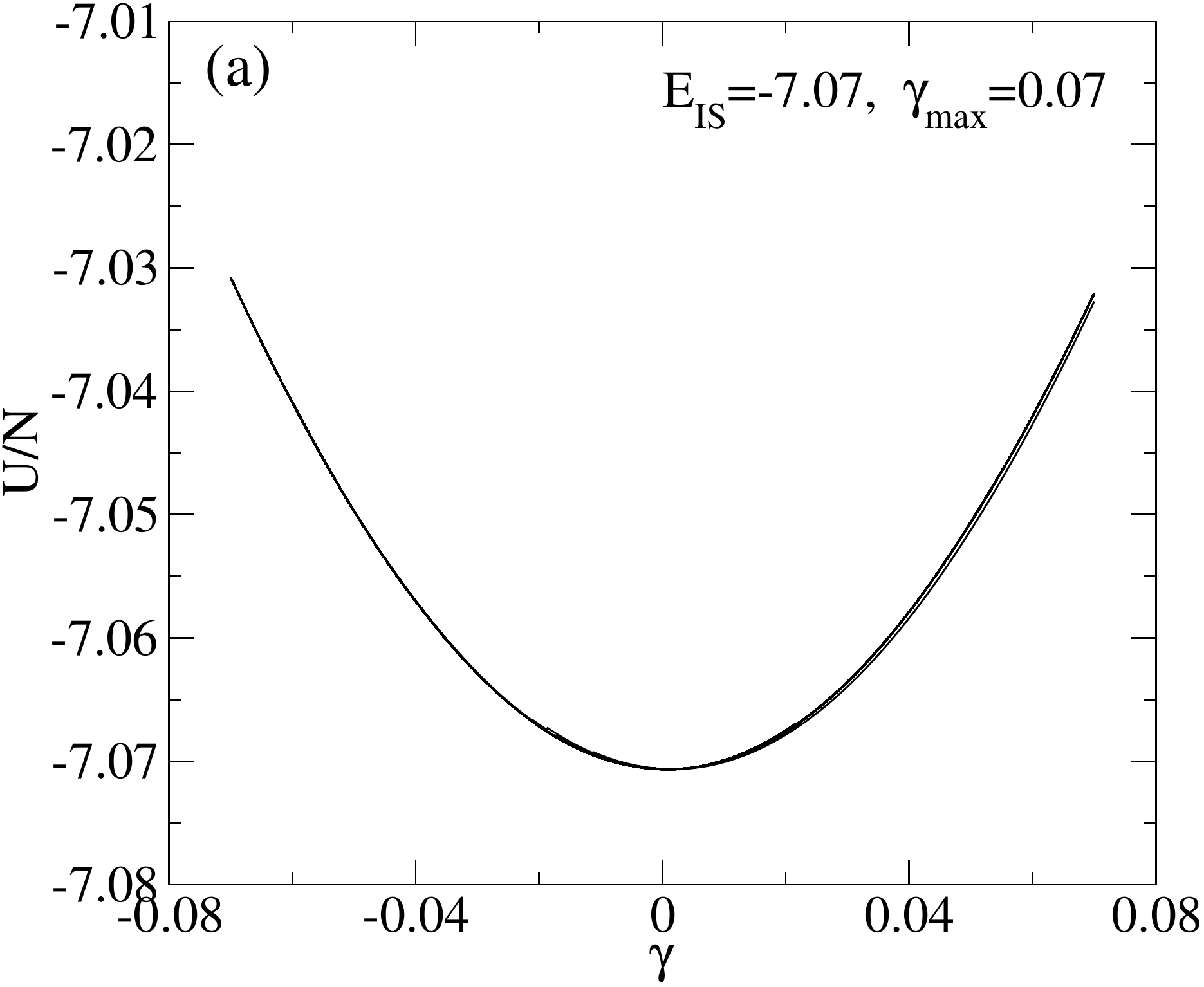}
  \includegraphics[width=.41\linewidth]{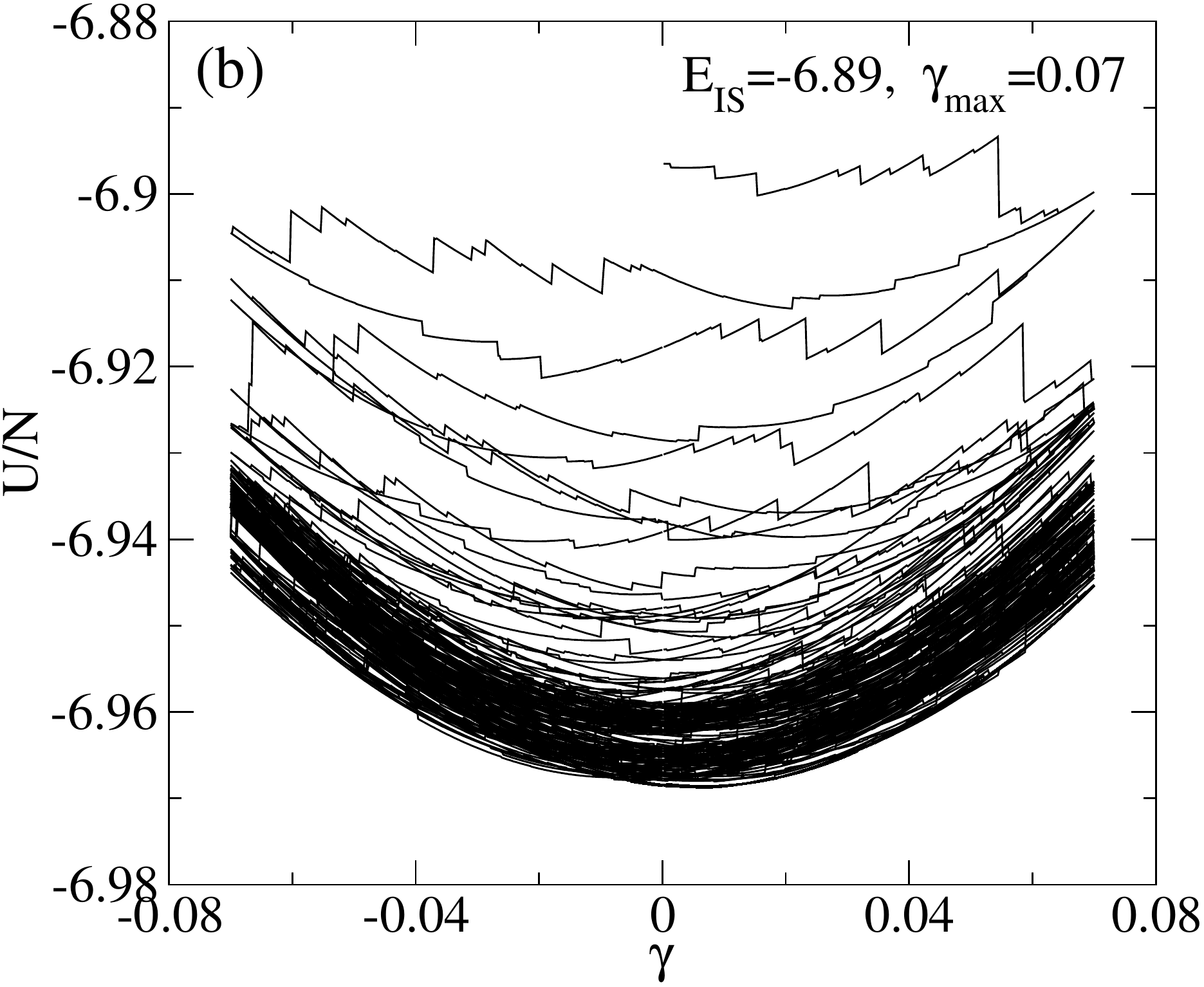}
     }
\caption{\label{SI_PE_cycle_BMLJ} {\bf Evolution of potential energy:}
  Starting from the initial states, the potential energy is shown as
  they approach the steady state, for different inherent structure
  energies (a) $E_{IS}=-7.07$ and (b) $E_{IS}=-6.89$, for a strain
  amplitude below yielding, $\gamma_{max} = 0.07$ Note that for the
  deeply annealed glass ($E_{IS}=-7.07$) all the cyclic curves lie on
  top of each other indicating that very little annealing takes
  place. For the poorly annealed glass ($E_{IS}=-6.89$), the system
  anneals, lowering the energy till a steady state is reached.}
\end{figure*}
In Fig. \ref{SI_PE_cycle_BMLJ}, we show the potential energy of the KA
BMLJ system for initial energies for strain amplitude
$\gamma_{max}=0.07$, below yielding. For the well annealed case
($E_{IS}=-7.07$), from the beginning of the strain cycles, the curves
traverse essentially the same energies, while for the poorly annealed
case $E_{IS}=-6.89$, starting from a high energy state, the system
anneals to a lower energy value till it reaches the steady state.

\clearpage
\subsection{Uniform shear deformation of KA BMLJ}

We have studied the deformation of the KA BMLJ system under uniform
shear. For such a study we consider a range of system sizes from
$N=2000$ to $64000$. We employ the AQS protocol with strain step
$d\gamma=0.0002$ and the strain varied from $0$ to $0.3$. For
$N=2000$, we considered $200$ samples, for $4000$ and $N=8000$, $100$
samples and for $32000$ and $64000$ we have considered $30-50$
samples.

\begin{figure*}[h]
\centerline{
\includegraphics[width=.32\linewidth]{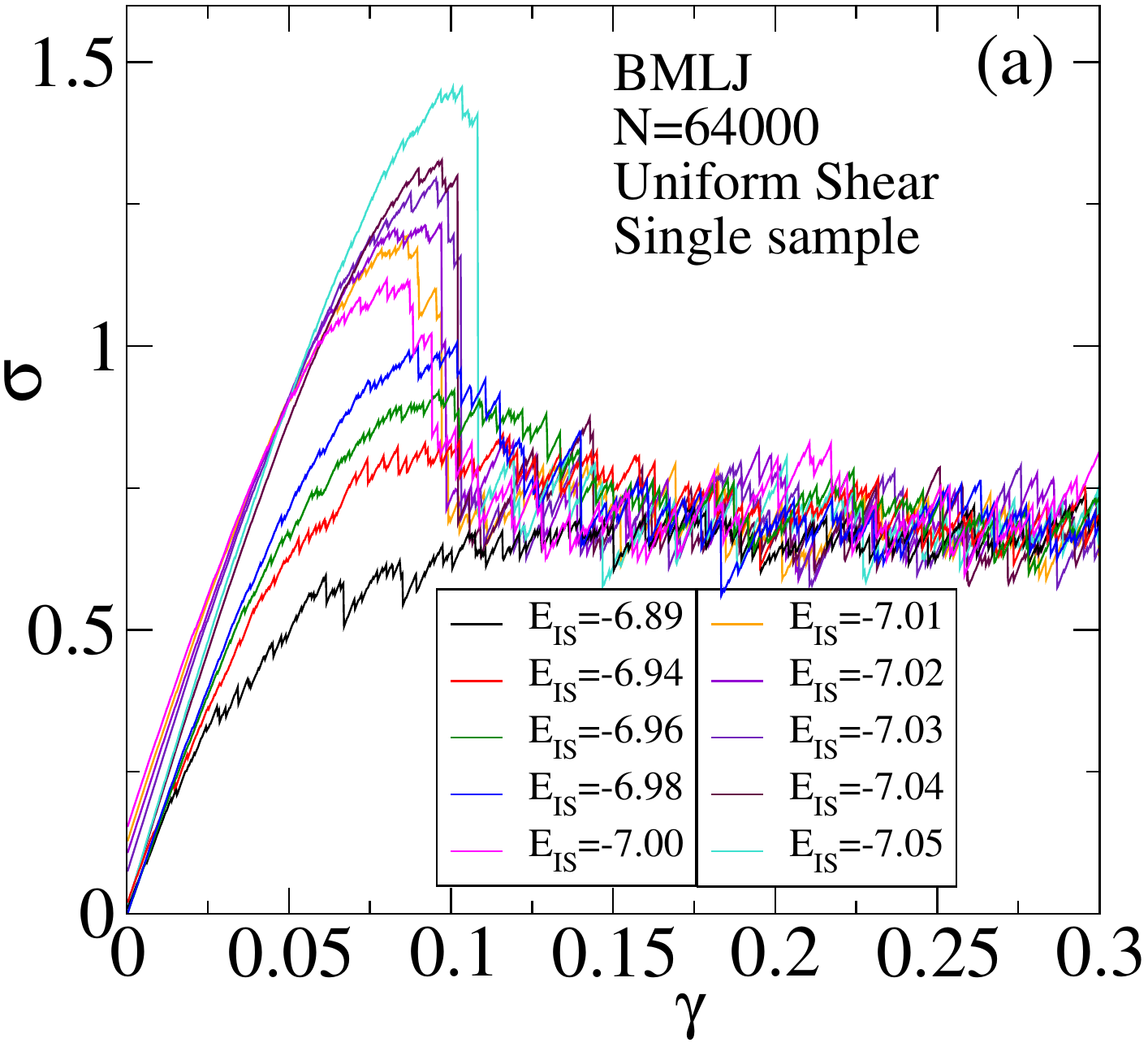}
\includegraphics[width=.32\linewidth]{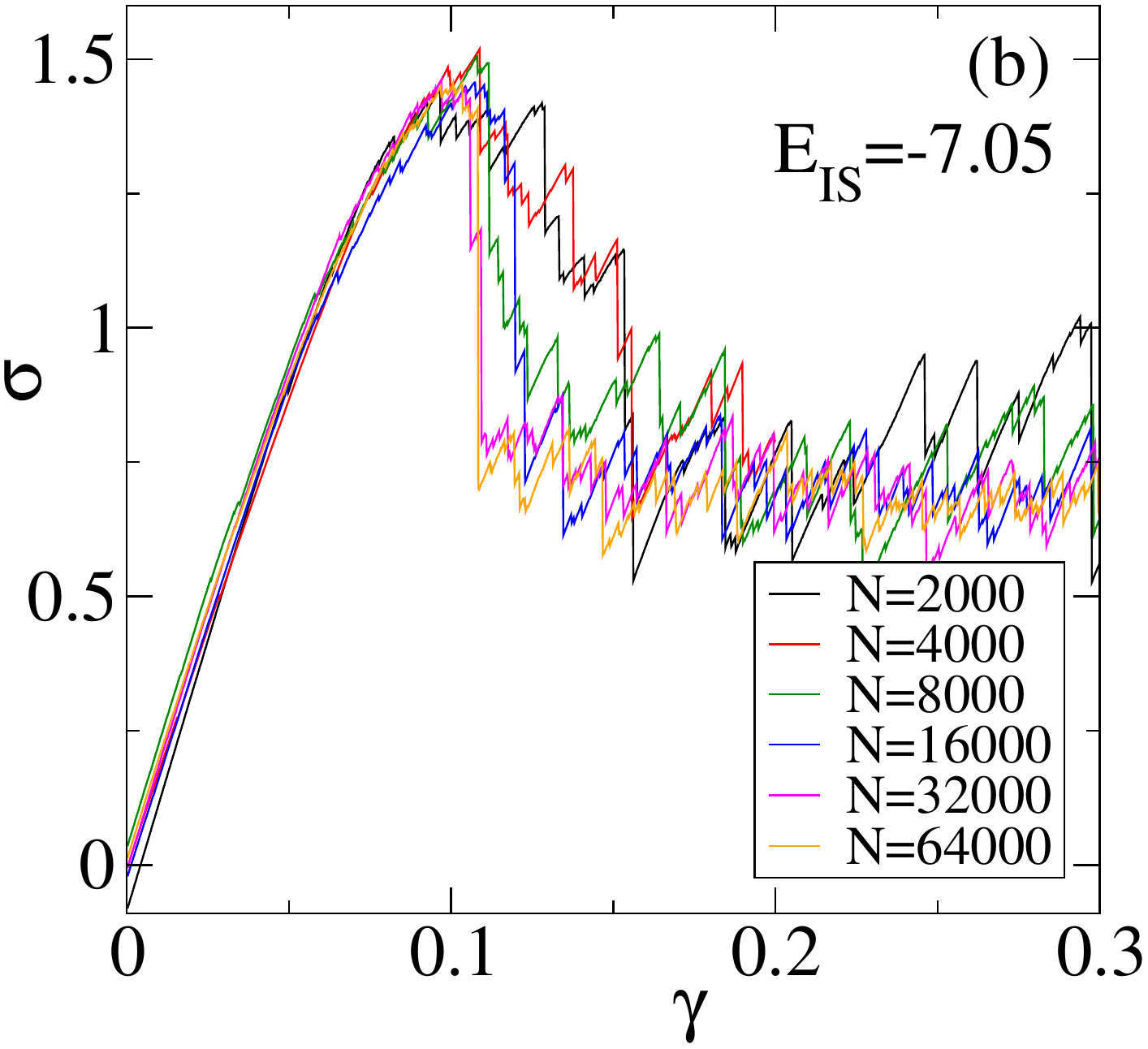}
    \includegraphics[width=.35\linewidth]{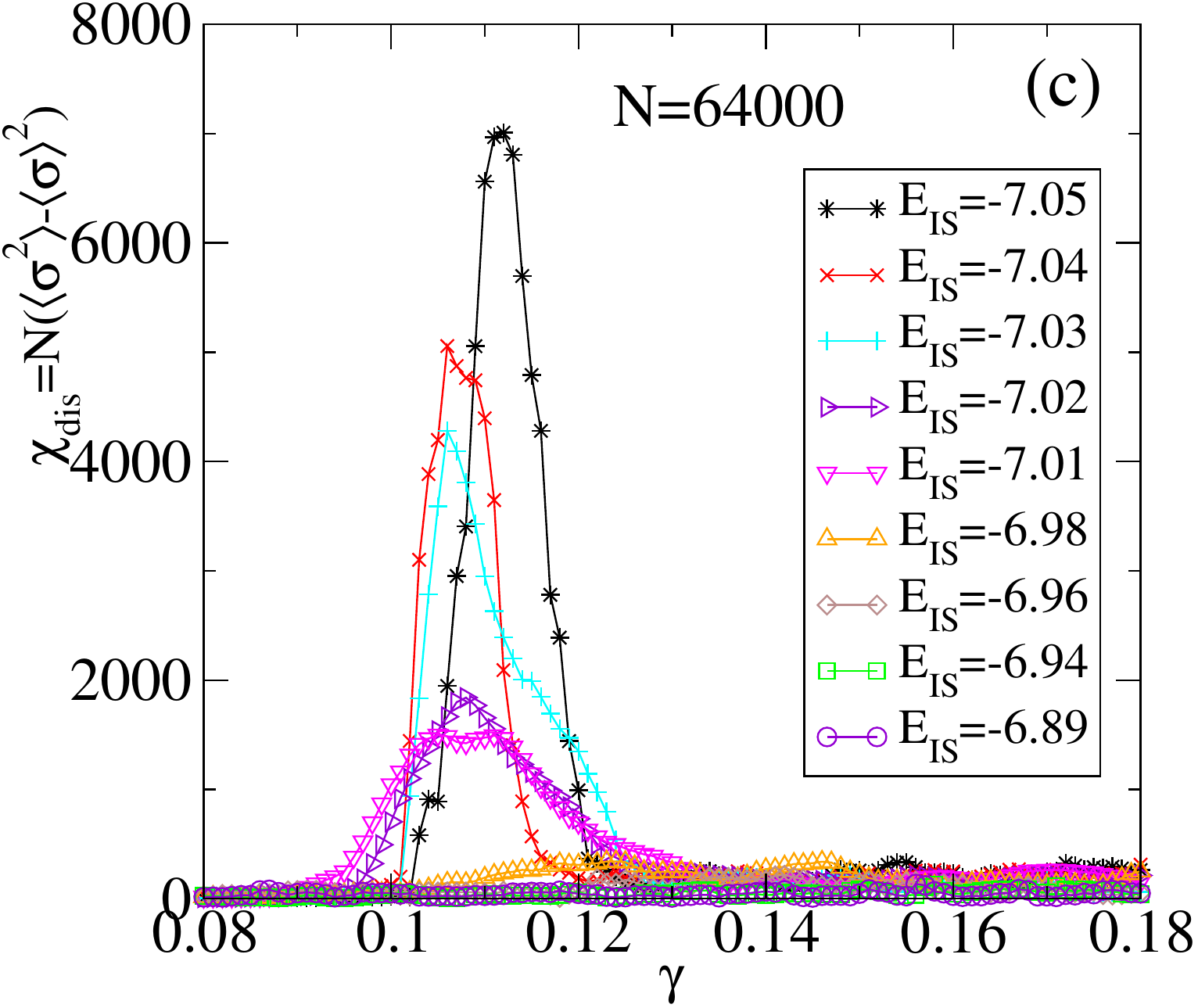}
    }
    \centerline{
\includegraphics[width=.34\linewidth]{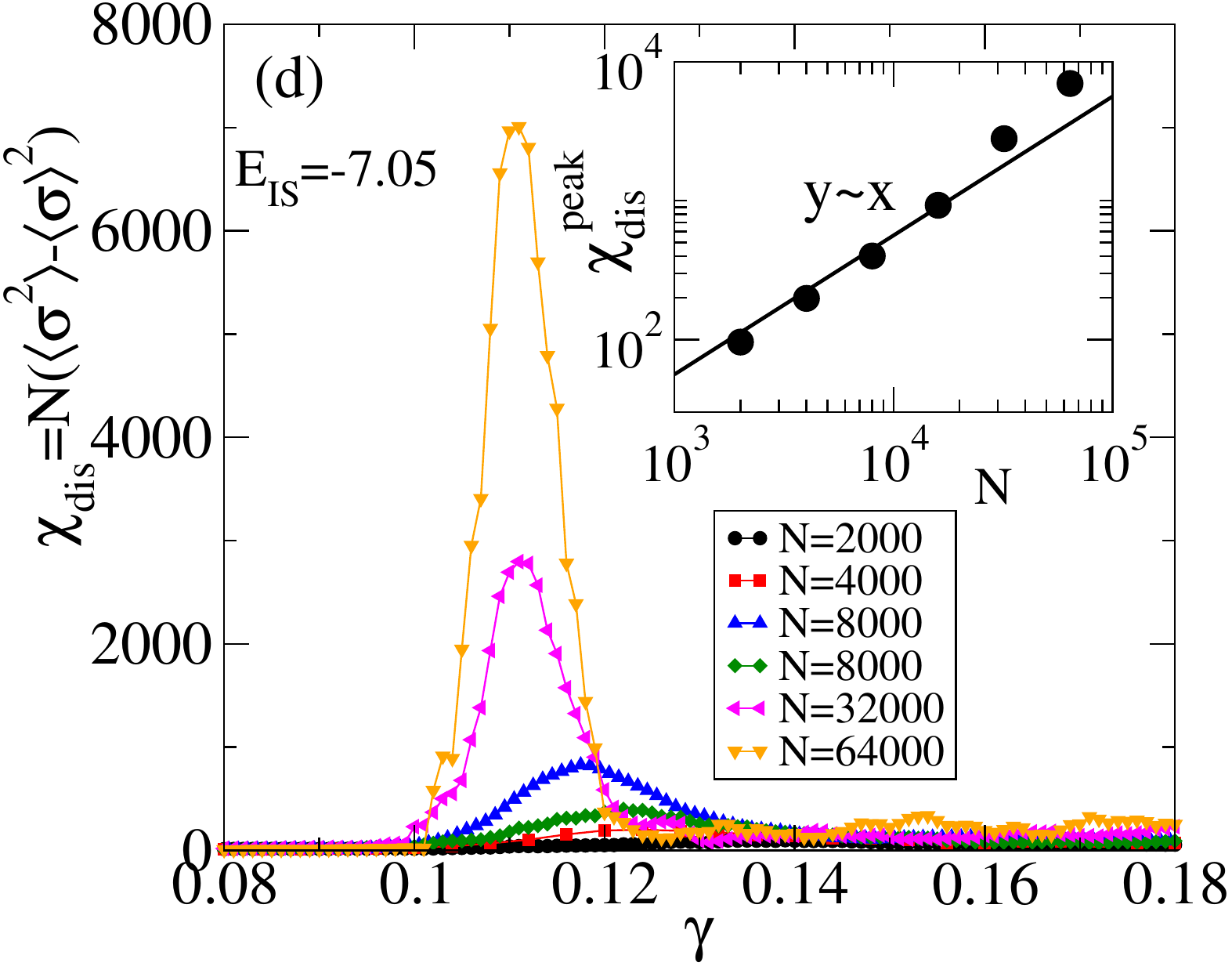}
  \includegraphics[width=.325\linewidth]{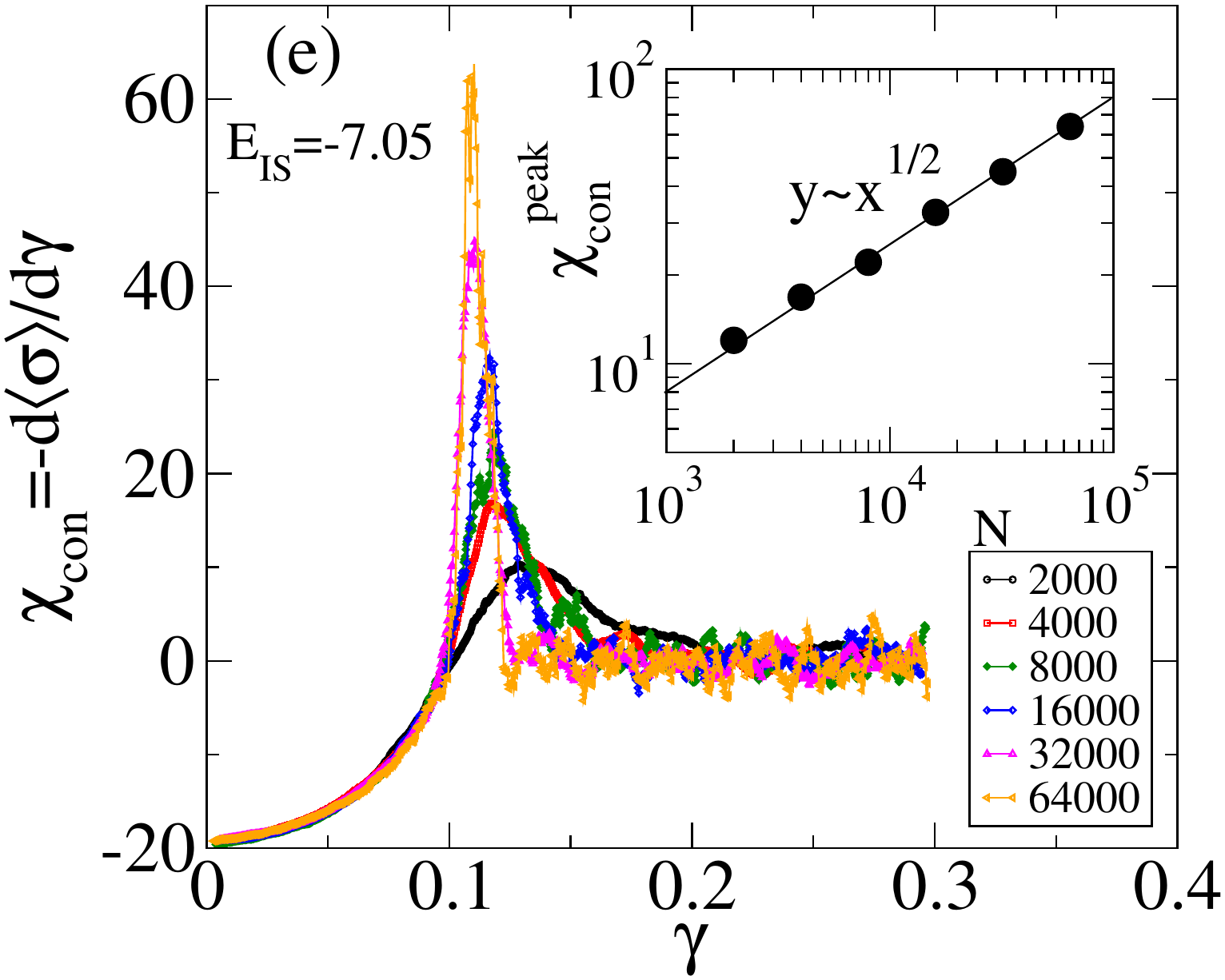}
    \includegraphics[width=.32\linewidth]{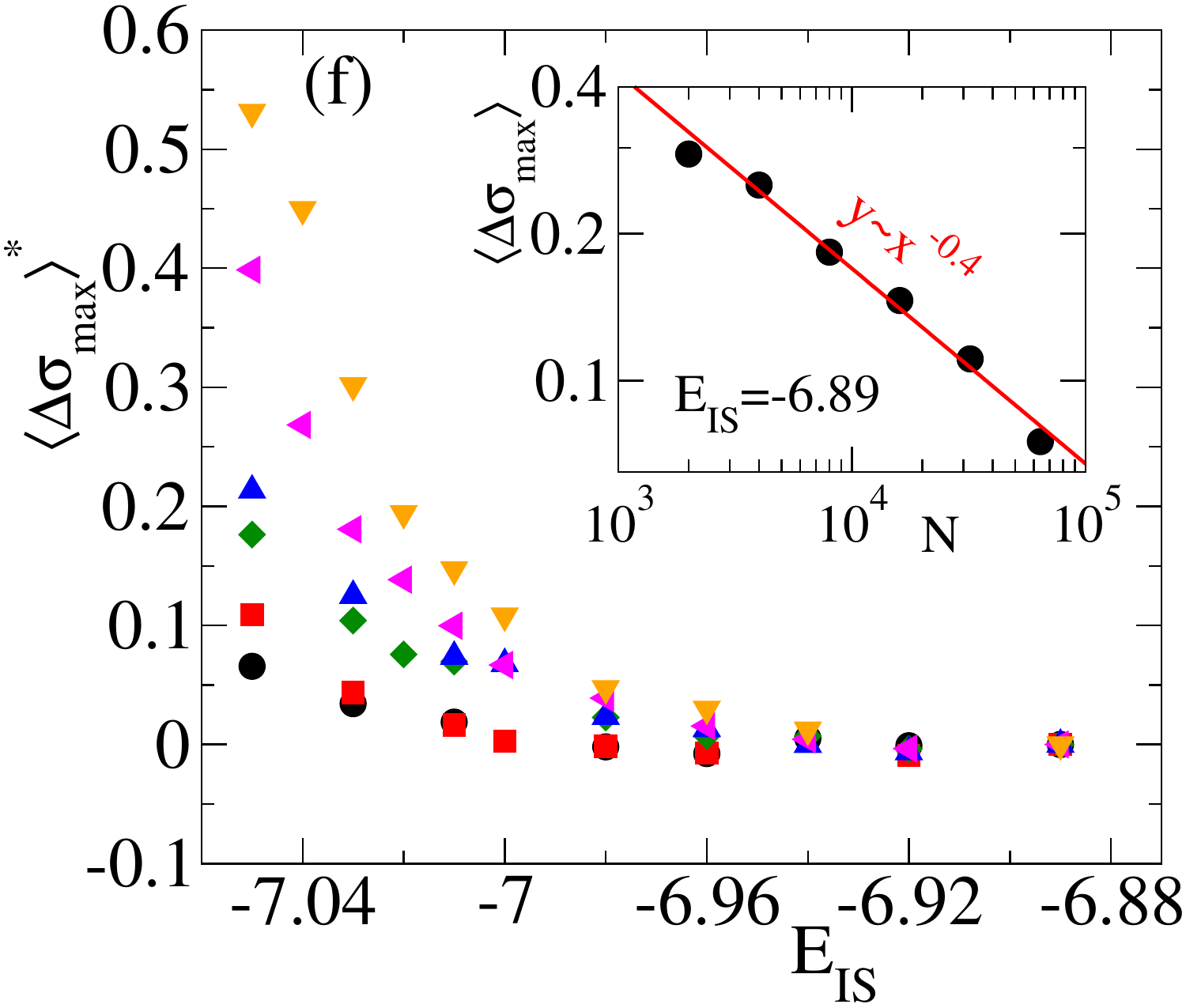}}
    \caption{\label{BMLJ_US} {\bf Uniform shear of KA BMLJ:} (a)
      Stress-strain curves for different $E_{IS}$ for a single sample,
      for the system of size $ N = 64000$. (b) Stress-strain curves
      for a single sample for $E_{IS}=-7.05$ for different system
      sizes. (c) Fluctuations in the stress (disconnected
      susceptibility) against $\gamma$ for different $E_{IS}$, $N =
      64000$. (d) disconnected and (e) connected susceptibility for
      different system sizes. Corresponding insets show the scaling of
      peak of susceptibility against $N$ showing
      $\chi_{dis}^{peak}\sim N$ and $\chi_{con}^{peak}\sim N^{1/2}$
      scaling, respectively.  (f) Plot of $\langle\Delta
      \sigma_{max}\rangle^{*} \equiv \langle\Delta \sigma_{max}\rangle
      -\langle\Delta \sigma_{max}\rangle_{E_{IS}=-6.89}$ , $\langle
      \Delta \sigma_{max}\rangle$ being the average maximum drop,
      against initial inherent structure for different system
      sizes. Inset: Plot of average maximum stress drop $\langle
      \Delta\sigma_{max}\rangle$ with system size $N$ for
      $E_{IS}=-6.89$.  }
\end{figure*}

In addition to data presented in the main text, here we show the
stress-strain curves for single samples for various cases. The
evolution of the shear stress under uniform shear deformation is shown
for different energies $E_{IS}$ for a single sample of size $N =
64000$ in Fig. \ref{BMLJ_US} (a). These curves reveal the emergence of
large single stress drops as the degree of annealing increases. The
stress response is shown in Fig. \ref{BMLJ_US} (b) for different
system sizes for a single sample of $E_{IS}$ ($=-7.05$) for each
size. These results illustrate the manner in which the stress drops
become focused in larger single drops at smaller strain values with an
increase in system size.

We compute the connected susceptibility, defined as
$\chi_{con}=-d\langle \sigma \rangle/ d\gamma$, considering the stress
averaged over $25$ ($40$ for lowest system size) samples, computing
mid-point numerical derivatives. The disconnected susceptibility,
$\chi_{dis} = N [\langle\sigma^2\rangle - \langle\sigma\rangle^2]$ is
obtained over the same sample set. We consider stress values within a
window of $\Delta \gamma=0.001$ for each sample, and obtain the mean
and variance over this data set at each average strain
value. Fig. \ref{BMLJ_US} (c) and (d) show $\chi_{dis}$ as a function
of $E_{IS}$ for $N = 64000$ and for $E_{IS} = -7.05$ for different
sizes respectively. The peak values in Fig. \ref{BMLJ_US} (d) are
shown in the inset as a function of system size. Fig. \ref{BMLJ_US}
(e) shows $\chi_{con}$ for $E_{IS} = -7.05$ for different sizes and
the peak values are shown in the inset as a function of system
size. Finally, in Fig.  \ref{BMLJ_US} (f), we show the dependence on
$E_{IS}$, for various system sizes, of the maximum stress drop
observed in a range of strain values from $0$ to $0.15$, straddling
the yield strain. We subtract the value of the stress drop at $E_{IS}
= -6.89$, which corresponds simply to the drops in the post-yield flow
regime. The maximum stress drop shows little system size dependence
above the threshold value of $E_{IS} = -6.99$ but grows strongly with
system size below, indicating the increasingly strong discontinuous
yielding at lower $E_{IS}$ or higher annealing.

\subsection{Intermediate scattering function}
\begin{figure*}[h]
\centerline{
    \includegraphics[width=.45\linewidth]{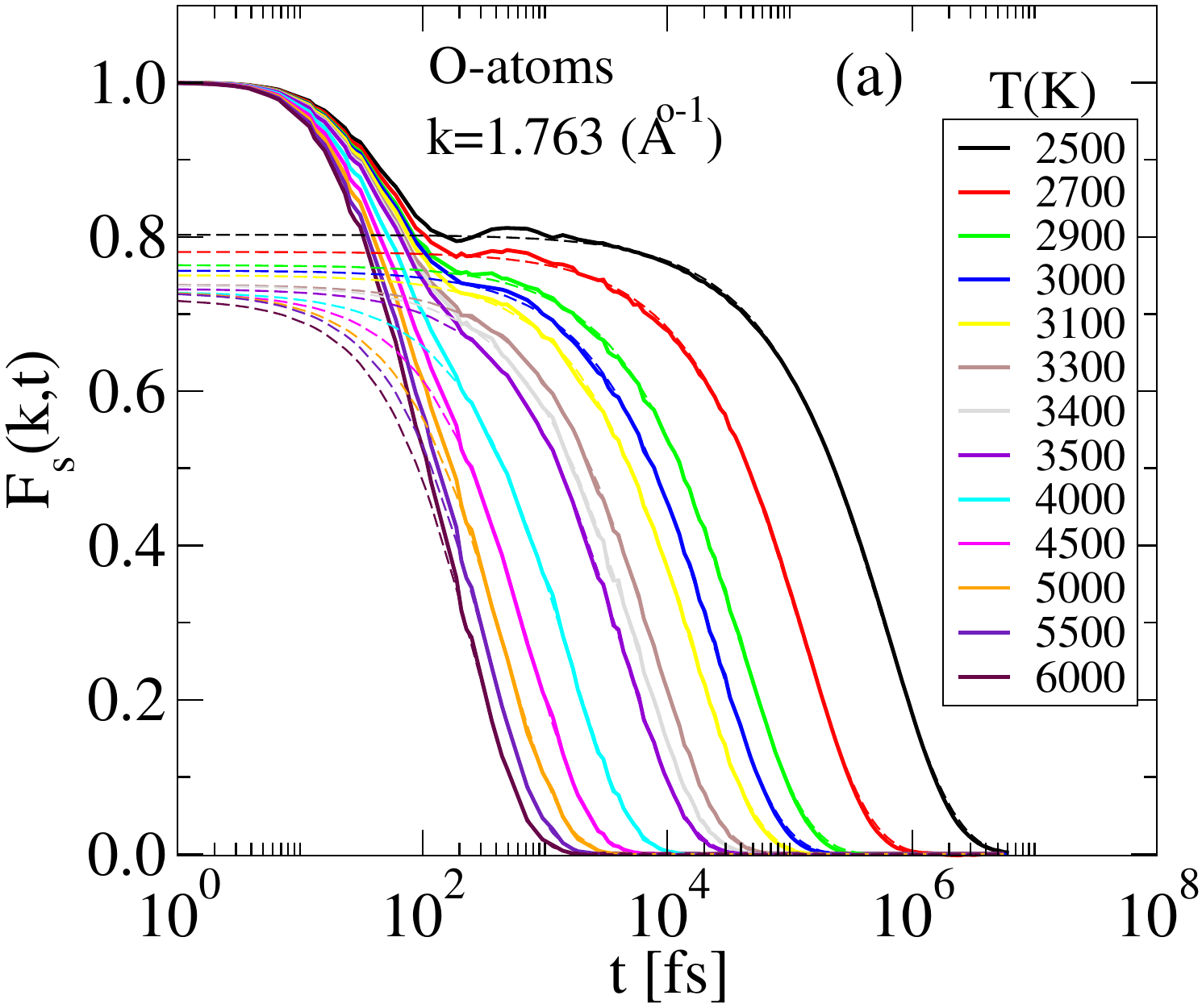}
    \includegraphics[width=.45\linewidth]{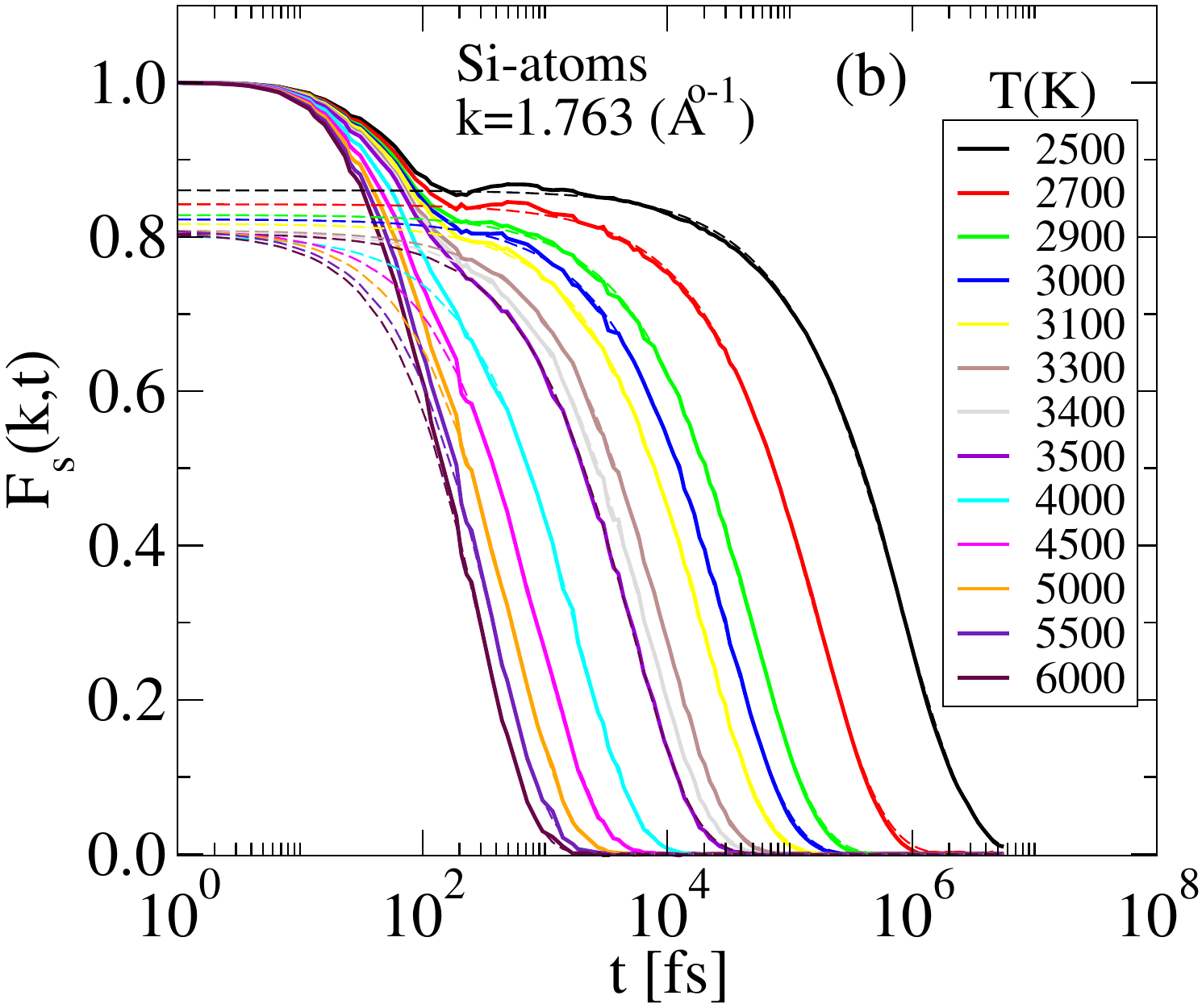}}
    \centerline{
     \includegraphics[width=.45\linewidth]{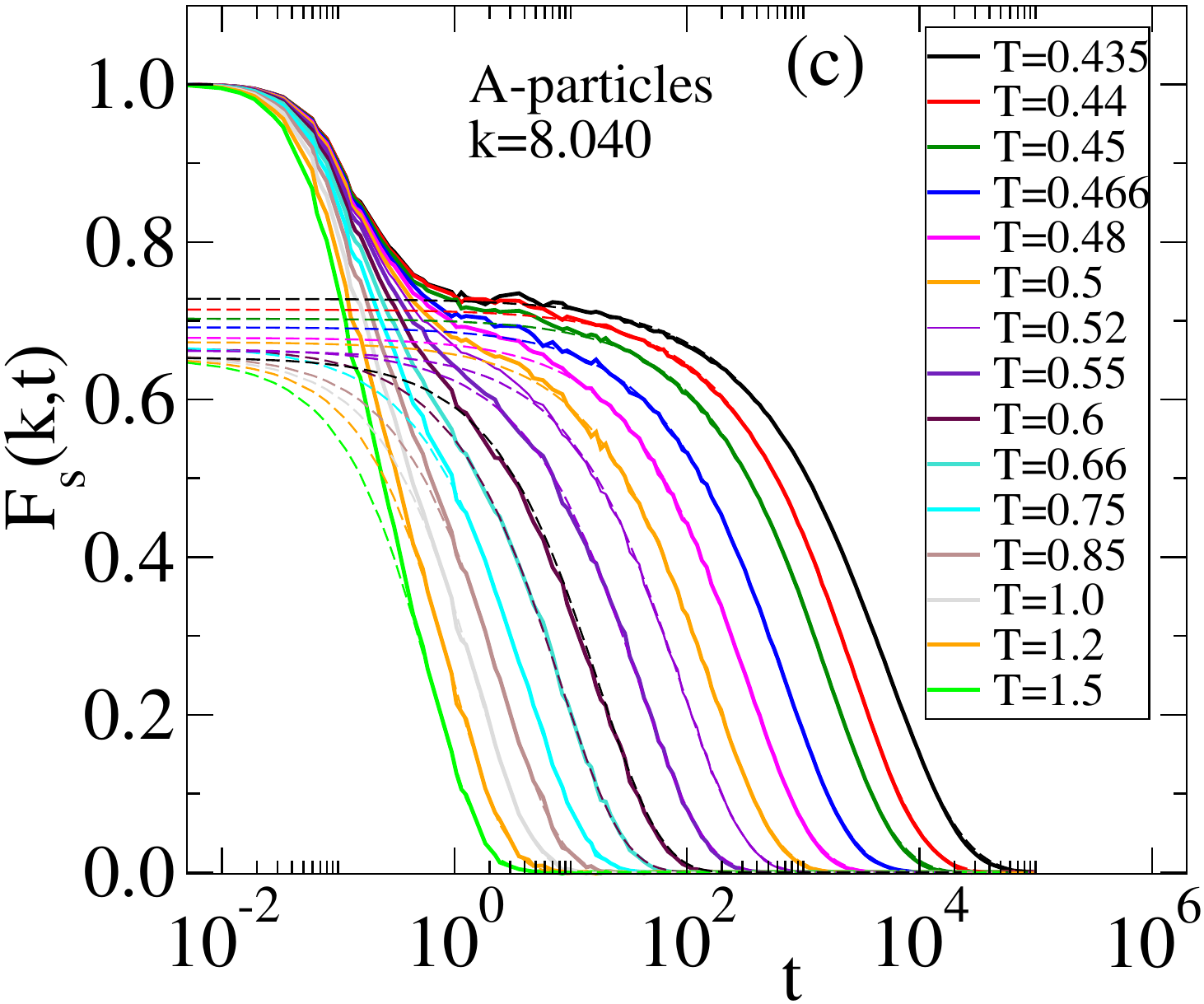}
     }
   \caption{\label{SI_fskt} {\bf Self intermediate scattering function
       of liquid silica and liquid BMLJ}: Plot of the self part of the
     intermediate scattering function $F_{s}(k,T)$ for (a) Oxygen and
     (b) Silicon atoms at different temperatures for a system of size
     $N=1728$ for $k=1.76 \AA^{-1}$. Dashed lines are fits to the form
     $f(t)=f_c[\exp(-(t/\tau_{\alpha})^\beta)]$. (c) Plot of
     $F_{s}(k,T)$ of A-type of particles for the KA BMLJ model for
     different temperatures. System size considered here is $N=4000$
     and $k=8.04$.}
\end{figure*}

In order to calculate the relaxation time of dynamics of liquid we have
calculated the self part of the intermediate scattering function $F_s(k,t)$. For A-type of particles (equivalently for Si and O atoms of silica) the $F_s(k,t)$ is defined as, \begin{equation}
F_s(k,t)=\frac{1}{N_A}\sum_{j=1}^{N_A}\langle \exp[i\boldsymbol{\rm
k}.(\boldsymbol{\rm r}_j(t)-\boldsymbol{\rm r}_j(0))]\rangle \nonumber
\end{equation}
where $k$ is the wave vector (with values used given the figure
caption), $r_i$ is the particle coordinate and $N_A$ is the number of
$A$-type atoms in the system. The calculated values of $F_s(k,t)$ are
shown in Fig. \ref{SI_fskt} for Oxygen and Silicon atoms for silica
and for the A-type particles for BMLJ for a wide range of
temperatures. The relaxation times $\tau_{\alpha}$ obtained from the
stretched exponential fits are shown in Fig. 4 of the main text.

\clearpage

\end{document}